\pdfoutput=1
\documentclass[aps,prl,10pt,twocolumn,notitlepage,noeprint,superscriptaddress]{revtex4-1}
\usepackage{amsmath,amssymb,amsfonts,graphicx,times,mathtools,amsthm,centernot}
\usepackage{bbm}
\usepackage{bbold}
\usepackage[utf8]{inputenc}
\usepackage{textcomp}
\usepackage{microtype}
\usepackage{bm}
\usepackage{braket}
\usepackage{xcolor}
\usepackage[breaklinks=true,colorlinks=true,linkcolor=blue,urlcolor=blue,citecolor=blue]{hyperref}
\newcommand{\id}{\mathbb{1}}
\newcommand{\zero}{\mathbb{0}}
\newcommand{\R}{\mathbb{R}}

\newcommand{\I}{\mathrm{i}}
\newcommand{\T}{\mathsf{T}}
\let\Re\relax
\let\Im\relax
\DeclareMathOperator{\Re}{Re}
\DeclareMathOperator{\Im}{Im}
\DeclareMathOperator{\Tr}{Tr}
\DeclareMathOperator{\tr}{Tr}
\DeclareMathOperator{\rnk}{rank}
\renewcommand{\vec}{\boldsymbol}
\let\originalleft\left
\let\originalright\right
\renewcommand{\left}{\mathopen{}\mathclose\bgroup\originalleft}
\renewcommand{\right}{\aftergroup\egroup\originalright}
\renewcommand{\right}{\aftergroup\egroup\originalright}

\begin{document}
\title{Evaluating the Holevo Cramér-Rao bound for multi-parameter quantum metrology}
\author{Francesco Albarelli}
\email{francesco.albarelli@gmail.com}
\affiliation{Department of Physics, University of Warwick, Coventry CV4 7AL, United Kingdom}
\author{Jamie F. Friel}
\email{j.friel@warwick.ac.uk}
\affiliation{Department of Physics, University of Warwick, Coventry CV4 7AL, United Kingdom}
\affiliation{EPSRC Centre for Doctoral Training in Diamond Science and Technology, UK}
\author{Animesh Datta}
\email{animesh.datta@warwick.ac.uk}
\affiliation{Department of Physics, University of Warwick, Coventry CV4 7AL, United Kingdom}
%%%
\date{\today}
%%%%
% However, its evaluation has remained elusive for problems of practical interest.
% , on the weighted sum of the covariance matrix elements of an estimator, 
% %For finite-dimensional systems we recast its evaluation as a semidefinite program, with reduced size for rank-deficient states, and use it to study phase and loss estimation in optical interferometry and three-dimensional magnetometry with noisy multi-qubit systems.
% Our finding for the former is that it is possible to attain the HCRB with single-copy measurements.
% For the latter, we show a non-trivial interplay between the HCRB and incompatibility, and bring evidence that single-copy projective measurements attain the HCRB in the noiseless two-qubit case.
\begin{abstract}
% The quantum aspects of metrology are fully revealed only with the simultaneous estimation of multiple parameters.
Only with the simultaneous estimation of multiple parameters are the quantum aspects of metrology fully revealed.
This is due to the incompatibility of observables.
The fundamental bound for multi-parameter quantum estimation is the Holevo Cram\'er-Rao bound (HCRB) whose evaluation has so far remained elusive.
For finite-dimensional systems we recast its evaluation as a semidefinite program, with reduced size for rank-deficient states.
We use this result to study phase and loss estimation in optical interferometry and three-dimensional magnetometry with noisy multi-qubit systems.
For the former, we show that, in some regimes, it is possible to attain the HCRB with the optimal (single-copy) measurement for phase estimation.
% single-copy measurements optimal for phase-estimation.
For the latter, we show a non-trivial interplay between the HCRB and incompatibility, and provide numerical evidence that projective single-copy measurements attain the HCRB in the noiseless two-qubit case.
\end{abstract}

\maketitle

\emph{Introduction}---Measuring physical quantities with ever increasing precision underlies both technological and scientific progress.
Quantum mechanics plays a central role in this challenge.
On the one hand, the unavoidable statistical uncertainty due to quantum fluctuations is a fundamental limitation to high precision metrology.
On the other hand, quantum-enhanced metrological schemes that take advantage of nonclassical features, such as entanglement, coherence, or squeezing, have been proposed and implemented experimentally~\cite{Giovannetti2011,Demkowicz-Dobrzanski2015a,Degen2016,Pezze2018,Braun2018,Pirandola2018,Berchera2018}.
Myriad metrological applications are intrinsically multi-parameter~\cite{Szczykulska2016}, e.g., sensing electric, magnetic or gravitational fields~\cite{Baumgratz2015}, force sensing~\cite{Tsang2011,Ang2013}, imaging~\cite{Humphreys2013,Gagatsos2016a}, superresolution~\cite{Tsang2016b,Chrostowski2017,Rehacek2017,Yu2018,Napoli2018}.
 % incoherent imaging microscopy and spectroscopy.
% A quantum-mechanical description of sensing schemes is vital in all these fields;
% For example, a quantum-mechanical analysis has recently led to new techniques for superresolution~\cite{Tsang2016b,Chrostowski2017,Rehacek2017,Tham2017,Parniak2018,Yu2018,Napoli2018}.
As a consequence, the field of multi-parameter quantum metrology has been growing rapidly, both theoretically~\cite{DAriano1997a,DAriano1998a,Macchiavello2003,Ballester2004a,Young2009,Vaneph2012,Genoni2013b,Zhang2014,Yuan2015,Berry2015a,Liu2017d,Liu2017h,Pezze2017,Eldredge2018,Nair2018,Nichols2018,Gessner2018,Kura2017,Yang2018b,Chen2019,Rubio2019} and experimentally~\cite{Vidrighin2014,Roccia2017,Roccia2018,Parniak2018,Polino2019}.

The mathematical framework behind quantum metrology is quantum estimation theory~\cite{Paris2009}, pioneered by Helstrom and Holevo~\cite{Helstrom1967,Helstrom1968,helstrom1976quantum,Holevo1973,Holevo1976,Holevo2011b}.
In particular, multi-parameter quantum estimation highlights a defining trait of quantum theory, absent in single parameter estimation: incompatibility of observables~\cite{Heinosaari2016,Zhu}.
Because of this, multi-parameter quantum estimation is much more challenging, but also serves as a testbed for understanding quantum measurements.

Precision bounds for multi-parameter estimation are given in terms of \emph{matrix} inequalities for the mean square error matrix (MSEM) $\vec{\Sigma}$, see Eq.~\eqref{eq:msem}. 
However, matrix bounds are in general not tight for multi-parameter quantum estimation.
Instead, the Holevo Cramér-Rao bound (HCRB)~\cite{Holevo2011b,Hayashi2005} is the most fundamental \emph{scalar} lower bound imposed by quantum mechanics on the weighted mean square error (WMSE) $\Tr \left[ W \vec{\Sigma} \right]$ (for a positive definite $W$).
The HCRB represents the best precision attainable with global measurements on an asymptotically large number of identical copies of a quantum state~\cite{Guta2006,Hayashi2008a,Kahn2009,Yamagata2013,Yang2018a}.
Implementing such collective measurements is exceptionally challenging~\cite{Roccia2017,Hou2018}, but in some cases the HCRB is attained by single-copy measurements: for pure states~\cite{Matsumoto2002} and for displacement estimation with Gaussian states~\cite{Holevo2011b}.

Despite its importance, the HCRB has been used more as a mathematical object in asymptotic quantum statistics~\cite{Gill2013} than applied to concrete metrological problems.
Indeed, the HCRB is considered hard to evaluate, even numerically, being defined through a constrained minimization over a set of operators.
Closed form results for non-trivial cases are known only for qubits~\cite{Suzuki2016a}, two-parameter estimation with pure states~\cite{[][{. Reprinted in Chapter 20 of Ref.~[54], an updated shorter version is published as Ref.~[61].}]Matsumoto1997}
%[54]->[Hayashi2005] and [61]->[Matsumoto2012]
and two-parameter displacement estimation with two-mode Gaussian states~\cite{Bradshaw2017a,Bradshaw2017}, while a numerical investigation has been attempted for pure states and Hamiltonian parameters~\cite{Gorecki2019}. 
The evaluation of the HCRB thus remains a major roadblock in the development of multi-parameter quantum metrology.

This Letter removes this roadblock by providing a recipe for evaluating the HCRB numerically for finite-dimensional systems.
Our main result recasts the optimization required for evaluating the HCRB as a semidefinite program (SDP).
This was shown only for displacement estimation with Gaussian states~\cite{Bradshaw2017}.
We present a SDP whose complexity grows with the rank of the state instead of a naïve dependence on the Hilbert space dimension.
% \AD{This is important since unlike }full state tomography~\cite{TeoBook2015,Artiles2005,YAMAGATA2011,Acharya2019}, typical problems in quantum metrology involve the estimation of \AD{a few parameters} with the highest possible precision.
The application of our recipe to evaluate the HCRB for two well-known metrological problems provides new insights.
In particular, we provide numerical evidence that single-copy attainability of the HCRB with projective measurements is possible in non-trivial cases.
% less taxing than previously believed.

\emph{Multi-parameter quantum estimation}---We consider a generic finite-dimensional quantum system with Hilbert space $\mathcal{H}\cong \mathbb{C}^d$, denote the space of linear operators ($d{\times}d$ matrices) on $\mathcal{H}$ as $\mathcal{L}\left(\mathcal{H}\right)\cong\mathbb{C}^{d{\times}d}$ and the space of observables (Hermitian matrices) as $\mathcal{L}_\mathsf{h}\left(\mathcal{H}\right)$.

The state of the system $\rho_{\vec{\theta}} \in \mathcal{L}_\mathsf{h}\left(\mathcal{H}\right)$ is parametrized by a real vector $(\theta_1, \dots, \theta_n)^\mathsf{T} = \vec{\theta} \in \Theta \subset \mathbb{R}^n$~\footnote{
We assume a sufficiently regular parametrization, in particular the state has a fixed rank for all $\vec{\theta}$.}, the collection $\{ \rho_{\vec{\theta}} \}$ for all the values of $\vec{\theta}$ is called the \emph{quantum statistical model}.
% of $n$ real parameters
The goal is to simultaneously estimate all $n$ parameters by measuring possibly multiple copies of $\rho_{\vec{\theta}}.$
After measurement, classical data is processed with an estimator $\tilde{\vec{\theta}}$, a function from the space of measurement outcomes $\Omega$ to the space of parameters $\Theta$.
The MSEM of the estimator
\begin{equation}
\label{eq:msem}
\vec{\Sigma}_{\vec{\theta}}\left(\vec{\Pi}, \tilde{\vec{\theta}} \right) = \sum_{\omega \in \Omega} p(\omega | \vec{\theta}) \left(\tilde{\vec{\theta}}(\omega) - \vec{\theta}\right)\left( \tilde{\vec{\theta}}(\omega) - \vec{\theta} \right)^\mathsf{T} ,
\end{equation}
quantifies the precision of the estimation.
The probability of observing the outcome $\omega$ is given by the Born rule $p(\omega | \vec{\theta})=\Tr(\rho_{\vec{\theta}} \Pi_\omega )$; the measurement is described by a positive operator valued measure (POVM): $\vec{\Pi}= \left\{ \Pi_\omega \succeq 0 , \omega \in \Omega \, \vert \, \sum_{\omega \in \Omega} \Pi_\omega = \id_d \right\}$, without loss of generality we consider $\Omega$ to be a finite set~\cite{Chiribella2007}.

We consider \emph{locally unbiased} estimators that satisfy
% unbiasedness conditions in a neighbourhood of the true value:
\begin{equation}
\label{eq:class_loc_unb}
\sum_{\omega \in \Omega} (\tilde{\theta}_i(\omega)-\theta_i) p(\omega| \vec{\theta} ) = 0 \, , \quad 
\sum_{\omega \in \Omega} \tilde{\theta}_i(\omega) \frac{\partial p(\omega| \vec{\theta} )}{\partial {\theta_j} } = \delta_{ij} \,.
\end{equation}
For this class of estimators the matrix Cram\'er-Rao bound (CRB) on the MSEM is~\cite{cramer1946mathematical}
\begin{equation}
\label{eq:CRB}
\vec{\Sigma}_{\vec{\theta}}\left(\vec{\Pi}, \tilde{\vec{\theta}} \right) \succeq F(\rho_{\vec{\theta}},\vec{\Pi})^{-1} \,
\end{equation}
($A \succeq 0$ iff $A$ is positive semidefinite); the classical Fisher information matrix (CFIM) $F(\rho_{\vec{\theta}},\vec{\Pi})$ is defined as
\begin{equation}
\label{eq:CFIM}
F(\rho_{\vec{\theta}},\vec{\Pi}) = \sum_{\omega \in \Omega} p( \omega | \vec{\theta} ) \left( \frac{ \partial \log p(\omega | \vec{\theta}) }{\partial \vec{\theta}} \right) \left( \frac{ \partial \log p(\omega | \vec{\theta}) }{\partial \vec{\theta}} \right)^\mathsf{T} ,
\end{equation}
where $\partial f(\vec{\theta}) / \partial \vec{\theta} $ is the gradient of the function $f$. 
% in what follows we always assume $F(\rho_{\vec{\theta}},\vec{\Pi})  \succ 0$.
For locally unbiased estimators the MSEM is the covariance matrix (CM) and the bound is attainable: there is always an estimator in this class with a CM equal to the inverse FIM~\cite{[][{. Reprinted as Chapter 8 in Ref.~[54].}]Nagaoka1989,Gill2013}.
%[54]->[Hayashi2005]
To meaningfully compare the precision of different multi-parameter estimators, it is customary to consider a \emph{scalar} cost function, the WMSE $\tr \left[ W \vec{\Sigma}_{\vec{\theta}}\left(\vec{\Pi}, \tilde{\vec{\theta}} \right) \right]$, with $0 \prec W \in \mathbb{S}^n $ ($\mathbb{S}^n$ is the set of real symmetric $n$-dimensional matrices).

Well-known bounds for multi-parameter quantum estimation are obtained from the symmetric logarithmic derivatives (SLDs) $L_i \in \mathcal{L}_\mathsf{h}(\mathcal{H})$, satisfying $ 2 \partial \rho_{\vec{\theta}} / \partial \theta_i = L_i \rho_{\vec{\theta}} + \rho_{\vec{\theta}}L_i $, and from the right logarithmic derivatives (RLDs) $\tilde{L}_i \in \mathcal{L}(\mathcal{H})$ satisfying $\partial \rho_{\vec{\theta}} / \partial \theta_i = \rho_{\vec{\theta}} \tilde{L}_i $.
The (real symmetric) quantum Fisher information matrix (QFIM) is defined as $J^\text{S}_{ij}=\Re\left( \Tr\left[ \rho_{\vec{\theta}} L_i L_j \right] \right)$~\cite{Helstrom1967,Helstrom1968}; an analogous (complex Hermitian) matrix in terms of the RLDs is $J^\text{R}_{ij}=\Tr\left[ \rho_{\vec{\theta}} \tilde{L}_i \tilde{L}_j \right]$~\cite{Yuen1973,Belavkin1976}.
Both matrices give lower bounds for the MSEM: $\vec{\Sigma}_{\vec{\theta}}\left(\vec{\Pi}, \tilde{\vec{\theta}} \right) \succeq \left(J^\text{S/R}\right)^{-1}$ and the corresponding scalar bounds are $C^\text{S}_{\vec{\theta}}(\rho_{\vec{\theta}};W)=\tr\left[W (J^\text{S})^{-1} \right]$ and $C^\text{R}_{\vec{\theta}}(\rho_{\vec{\theta}};W)=\tr\left[W \Re(J^\text{R})^{-1} \right] + \left\Vert \sqrt{W} \Im (J^\text{R})^{-1} \sqrt{W} \right\Vert_1 $, where $\Vert A \Vert_1 = \tr\left[ \sqrt{A^\dag A} \right]$ is the trace norm~\cite{Holevo2011b,Nagaoka1989,Hayashi2017c}.

Holevo introduced a tighter bound, the HCRB $C^{\text{H}}_{\vec{\theta}}$~\cite{Holevo1976,Holevo2011b}:
\begin{equation}
\tr \left[ W \vec{\Sigma}_{\vec{\theta}}\left(\vec{\Pi}, \tilde{\vec{\theta}} \right) \right] \geq C^{\text{H}}_{\vec{\theta}} \geq  \max \left\{ C^{\text{S}}_{\vec{\theta}} , C^{\text{R}}_{\vec{\theta}} \right\}.
\end{equation}
Both inequalities can be tight.
In particular~\cite{Ragy2016}:
\begin{equation}
\label{eq:weak_comm}
C^{\text{H}}_{\vec{\theta}}(\rho_{\vec{\theta}};W)=C^{\text{S}}_{\vec{\theta}}(\rho_{\vec{\theta}};W) 
% \quad \forall \, \mathbb{S}^n \ni W \succ 0 
\iff 
D_{\vec{\theta}} = \zero_{n},
\end{equation}
where $(D_{\vec{\theta}})_{ij} \equiv \Im \left( \Tr\left[ L_j L_i \rho_{\vec{\theta}}\right] \right)$ is a skew-symmetric matrix~\cite{Nagaoka1989}.
Condition~\eqref{eq:weak_comm} is called \emph{weak commutativity} and quantum statistical models satisfying it \emph{asymptotically classical}~\cite{Suzuki2018}.

% In particular, $C^{\text{H}}_{\vec{\theta}}(\rho_{\vec{\theta}};W)=C^{\text{S}}_{\vec{\theta}}(\rho_{\vec{\theta}};W) \quad \forall \, \mathbb{S}^n \ni W \succ 0$ if and only if the following \emph{weak commutativity} condition holds $\forall i,j$~\cite{Ragy2016}:
% \begin{equation}
% \label{eq:weak_comm}
% (D_{\vec{\theta}})_{ij} \equiv \Im \left( \Tr\left[ L_i L_j \rho_{\vec{\theta}}\right] \right) = \frac{\I} \Tr\left[ \left[ L_i , L_j \right] \rho_{\vec{\theta}} \right]=0 
% \end{equation}
% Quantum statistical models satisfying this condition have been labeled \emph{asymptotically classical}~\cite{Suzuki2018}.
% , or \emph{compatible}~\cite{Ragy2016}.

\emph{Computing the bound with a SDP}---The HCRB is obtained as the result of the following minimization~\cite{Holevo2011b,Hayashi2017c}:
\begin{equation}
\label{eq:HCRB}
C_{\vec{\theta}}^\text{H} \left( \rho_{\vec{\theta}} ; W \right) = \min_{V \in \mathbb{S}^n , \vec{X} \in \mathcal{X}_{\vec{\theta}} } \left( \tr \left[ W V \right] \; \vert \; V \succeq Z \left[ \vec{X} \right] \right) \,,
\end{equation}
with the Hermitian $n{\times}n$ matrix $Z \left[ \vec{X} \right]_{ij}=\Tr \left[ X_i X_j \rho_{\vec{\theta}} \right]$ and the collection $\vec{X}$ of operators $X_i \in \mathcal{L}_\mathsf{h}(\mathcal{H})$ in the set
\begin{equation}
\mathcal{X}_{\vec{\theta}} = \left\{   \vec{X}   = \left( X_1, \dots, X_n \right) \, \vert  \, \Tr \left[ X_i \partial_j \rho_{\vec{\theta}} \right]   =\delta_{ij} \right\} \,.
\end{equation}
% \begin{equation}
% \begin{split}
% \mathcal{X}_{\vec{\theta}} = & \left\{   \vec{X}   = \left( X_1, \dots, X_n \right)  \vert \, X_i \in \mathcal{L}_\mathsf{h}(\mathcal{H}), \right. \\ 
% & \; \left. \Tr\left[ X_i \rho_{\vec{\theta}} \right]= 0 , \, \Tr \left[ X_i \partial_j \rho_{\vec{\theta}} \right]   =\delta_{ij} \right\} \,;
% \end{split}
% \end{equation}
% to enforce local unbiasedness.
% ~\eqref{eq:class_loc_unb} in the quantum case.
% The condition $\Tr\left[ X_i \rho_{\vec{\theta}} \right]= 0$ can be dropped, since it does not change the result, see e.g.~\cite[Ex. 6.57]{Hayashi2017c}.

For a density matrix with rank $r < d$ we can restrict the operators $X_i$ to the quotient space $\mathcal{L}^r_\mathsf{h}(\mathcal{H}) = \mathcal{L}_\mathsf{h}(\mathcal{H}) / \mathcal{L}_\mathsf{h}\left(\mathrm{ker}(\rho_{\vec{\theta}})\right)$, with dimension $\tilde{d}=2d r - r^2$.
For any $X \in \mathcal{L}_\mathsf{h}(\mathcal{H}),$ any scalar quantity evaluated in the eigenbasis of $\rho_{\vec{\theta}}$ is independent of the diagonal block of $X$ corresponding to the kernel of $\rho_{\vec{\theta}}$~\cite{Holevo2011b,Fujiwara1995} (see Sec.~I of~\cite{SM} for details).
% More details are given in the supplementary material (SM).

We introduce a basis $\lambda_i$ of Hermitian operators for $\mathcal{L}_\mathsf{h}^r(\mathcal{H})$, orthonormal w.r.t. the Hilbert-Schmidt inner product $\Tr \left[ \lambda_i \lambda_j \right] = \delta_{ij}$.
Using such a basis, each operator $X_i \in \mathcal{L}^r_\mathsf{h}(\mathcal{H})$ corresponds to a real valued vector $\vec{x}_i \in \R^{\tilde{d}}$.
With some abuse of notation we use $\vec{X}$ to denote also the collection of these real vectors, i.e., the $\tilde{d}{\times}n$ real matrix with $\vec{x}_i$ as columns.
The quantum state also belongs to $\mathcal{L}^{r}_\mathsf{h}(\mathcal{H})$ and therefore corresponds to a vector $\vec{s}_{\vec{\theta}}$ in the chosen basis.
This corresponds to the generalized Bloch vector~\cite{Bertlmann2008,Watanabe2011} when working in the full space $\mathcal{L}_\mathsf{h}(\mathcal{H})$.

A quantum state induces an inner product on $\mathcal{L}^{r}_\mathsf{h}\left(\mathcal{H}\right)$ via
\begin{equation}
\label{eq:RLDinner}
Z \left[ \vec{X} \right]_{ij} = \Tr \left[ X_i X_j \rho_{\vec{\theta}} \right] = \vec{x}_i^\mathsf{T} S_{\vec{\theta}} \vec{x}_j,
\end{equation}
where $S_{\vec{\theta}} \succeq 0 $ is the Hermitian matrix representing the inner product in the chosen basis.
With this choice we can write $Z\left[ \vec{X} \right] = \vec{X}^{\mathsf{T}} S_{\vec{\theta}} \vec{X}$ so that the matrix inequality on the r.h.s of Eq.~\eqref{eq:HCRB} reads $V \succeq \vec{X}^{\mathsf{T}} S_{\vec{\theta}} \vec{X}$.
Crucially, this last matrix inequality can be converted to a \emph{linear} matrix inequality (LMI) by using the Schur complement condition for positive semidefiniteness~\cite{Zhang2005}:
\begin{equation}
\label{eq:SchurPosDef}
V - B^\dag B \succeq 0 \iff \begin{pmatrix} V & B^\dag \\ B & \id  \end{pmatrix} \succeq 0 ,
\end{equation}
for any matrix $B$ and identity matrix $\id$ of appropriate size.
Thus, we can rewrite the minimization problem in Eq.~\eqref{eq:HCRB} as
\begin{equation}
\label{eq:SDPform2}
\begin{aligned}
& \underset{V \in \mathbb{S}^{n},\vec{X} \in \mathbb{R}^{\tilde{d}{\times}n} }{\text{minimize}} & & \tr \left[W  V  \right]\\
& \quad \text{subject to} & & \begin{pmatrix} V & \vec{X}^\mathsf{T} R_{\vec{\theta}}^\dag \\
R_{\vec{\theta}} \vec{X} & \id_{\tilde{r}} \end{pmatrix} \succeq 0 \\ 
& & & \vec{X}^\mathsf{T} \, \frac{\partial \vec{s}_{\vec{\theta}}}{\partial \vec{\theta} } = \id_{n} \,
\end{aligned}\;,
\end{equation}
where the matrix $R_{\vec{\theta}}$ can be \emph{any} $\tilde{r}{\times}\tilde{d}$ matrix (with $r d = \rnk\left( S_{\vec{\theta}} \right) \leq \tilde{r} \leq \tilde{d}$) satisfying $S_{\vec{\theta}} = R_{\vec{\theta}}^\dag R_{\vec{\theta}} $, e.g. a Cholesky-like decomposition.
Here $\partial \vec{s}_{\vec{\theta}}/\partial \vec{\theta}$ is a matrix with the vector components of the operators $\partial \rho_{\vec{\theta}} / \partial \theta_i = \partial_i \rho_{\vec{\theta}} $ as columns; this is the Jacobian matrix of $s_{\vec{\theta}}$ only if the basis $\left\{ \lambda_i \right\}$ is parameter independent.
The program~\eqref{eq:SDPform2} can be readily recognized as a convex minimization problem~\cite{Boyd}, since the set of solutions to LMI is convex and the objective function is linear.
It can be converted to a SDP (see Sec.~II of~\cite{SM} for details),
% (see SM for more details),
which can be solved numerically using efficient and readily available algorithms with a guarantee of global optimality.
In practice, the program~\eqref{eq:SDPform2} can be fed directly to a numerical modeling framework, such as \textsc{CVX}~\cite{cvx} or \textsc{YALMIP}~\cite{Lofberg2004}.
% , that convert it to a SDP in a standard form accepted by numerical solvers.

This convex optimization satisfies Slater's condition for strong duality~\cite{Boyd}, as long as $J^\text{S} \succ 0$, i.e. a non-singular quantum statistical model.
Strong duality holds when the minimum is finite and coincides with the maximum of the Lagrangian dual problem.
Qualitatively, Slater's condition means that the interior of the set of feasible points for~\eqref{eq:SDPform2} must not be empty.
We denote by $\vec{L}$ the matrix with the real vectors representing the SLDs as columns.
Upon noticing that $( \vec{L}^\T \partial \vec{s}_{\vec{\theta}}/ \partial \vec{\theta} )_{ij}= \Tr\left[ L_i \partial_j \rho_{\vec{\theta}} \right] = (J^{\text{S}})_{ij}$ it is easy to show that the matrices $\vec{X} = \vec{L} \, (J^{\text{S}})^{-1}$ and $V=(J^\text{S})^{-1} + V'$, with an arbitrary $V'\succ 0$, satisfy both constraints in~\eqref{eq:SDPform2}. 
For this choice of $V$ and $\vec{X}$ the matrix inequality in~\eqref{eq:HCRB} and~\eqref{eq:SDPform2} are strict and Slater's condition is satisfied.

An analytical optimization over $V$ in ~\eqref{eq:HCRB} leads to~\cite{Nagaoka1989}: 
\begin{align}
h_{\vec{\theta}}(\vec{X}) = &\min_{V \in \mathbb{S}^n } \left( \tr \left[ W V \right] \; \vert \; V \succeq Z \left[ \vec{X} \right] \right) = \label{eq:HolFunImpl}\\
= & \tr \left[ W \mathrm{Re} Z[\vec{X}] \right] +   \left\Vert \sqrt{W} \mathrm{Im} Z[\vec{X}] \sqrt{W} \right\Vert_1 \,, \label{eq:HolFunExpl}
\end{align}
so that $C_{\vec{\theta}}^\text{H} \left( \rho_{\vec{\theta}} ; W \right) = \min_{\vec{X} \in \mathcal{X}_{\vec{\theta}}} h_{\vec{\theta}}(\vec{X})  $ and usually this last optimization is solved numerically.
From our previous convexity argument we also infer that $h_{\vec{\theta}}(\vec{X}) $ is a convex function of $\vec{X}$; being a partial minimization of an affine function over a convex set~\cite{Boyd}.
This may not be apparent from~\eqref{eq:HolFunExpl} since the second term is not convex; the sum of the two terms is convex as long as the matrix $Z[\vec{X}]$ is positive semidefinite and the identity~\eqref{eq:SchurPosDef} can be used.

\emph{Optical interferometry with loss}---Optical interferometry, where the goal is to measure a phase difference between two optical paths, is a prime example of quantum metrology~\cite{Demkowicz-Dobrzanski2015a}. 
In some instances, one may wish to estimate both the phase and the loss induced by a sample in one arm of a Mach-Zehnder interferometer~\cite{Crowley2014}.
% For instance, one may wish to estimate both the phase and the loss induced by a sample in one arm of a Mach-Zehnder interferometer~\cite{Crowley2014} or to estimate the intrinsic loss of the interferometer, whose knowledge is needed for the preparation of optimal protocols~\cite{Demkowicz-Dobrzanski2009,Dorner2009,Demkowicz-Dobrzanski2012,Lee2009,Dinani2014}.

We consider initial states with a fixed photon number $N$ across two modes $| \psi_\text{in} \rangle = \sum_{k=0}^{N} c_k | k ,  N-k \rangle.$
These include for example N00N states and Holland-Burnett states~\cite{Holland1993}.
The evolved state after the lossy interferometer, with one arm characterized by a transmissivity $\eta$ and a phase shift $\phi$, has a direct sum form $\rho_{\phi , \eta} = \bigoplus_{k=0}^N p_l | \psi_l \rangle \langle \psi_l |$, where each $|\psi_l \rangle$ corresponds to $l$ lost photons~\cite{Demkowicz-Dobrzanski2009} (see Sec.~III of~\cite{SM} for details).
For this problem it is possible to obtain the SLDs $L_\phi$ and $L_\eta$ analytically, as well as the QFIM $J^\text{S}=\mathrm{diag}(J^\text{S}_{\phi \phi},J^\text{S}_{\eta \eta})$.
Crucially, this multi-parameter estimation problem is \emph{never} asymptotically classical, since~\cite{Crowley2014}
\begin{equation}
\label{eq:PhaseLossNoAsClass}
\Im \left( \Tr\left[ L_\phi L_\eta \rho_{\phi,\eta} \right] \right) = - \frac{J^\text{S}_{\phi \phi}}{2 \eta} \,.
\end{equation}
Hence the weak commutativity condition~\eqref{eq:weak_comm} never holds if the model is non-singular; thus we get $C_{\vec{\theta}}^\text{H} > C_{\vec{\theta}}^\text{S} > C_{\vec{\theta}}^\text{R}  = 0$ (the RLD bound is completely uninformative~\cite{Crowley2014}).
Eq.~\eqref{eq:PhaseLossNoAsClass} also means that phase and loss cannot be jointly estimated with the same precision obtainable by estimating each parameter individually and there exists a trade-off between precisions.
% in joint phase and loss estimation.
Following~\citet{Crowley2014} we focus on a strategy to estimate $\phi$ with the best possible precision and still get an estimate of $\eta$, by considering the projective POVM $\vec{\Pi}_\phi$ obtained from the spectral decomposition of the SLD $L_\phi$.

\begin{figure}[b]
\centering
\includegraphics{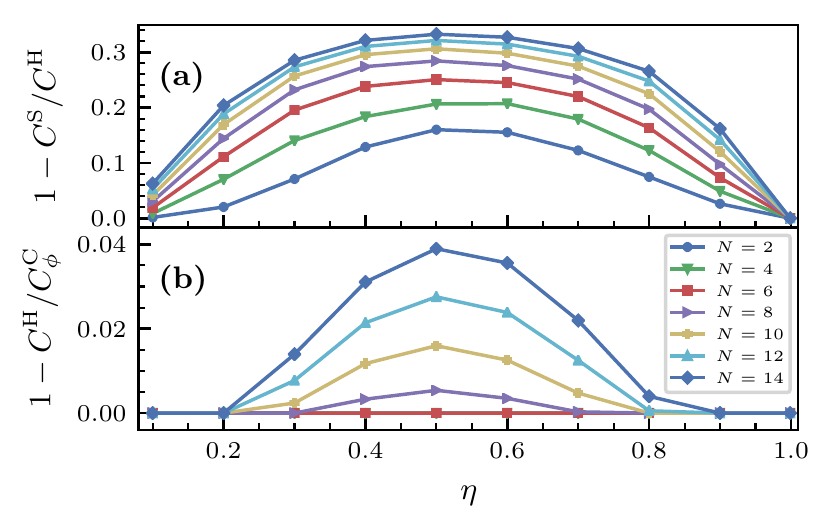}
\caption{Relative difference between different CRBs for simultaneous estimation of phase and loss, as a function of the transmissivity $\eta$, for $N$-photons Holland-Burnett probe states and $W=\id_2$.
% \textbf{(a)} The asterisk marks represent the HCRB $C^{\text{H}}_{\phi ,\eta}$, the solid lines represent the classical CRB $C^{\text{C}}_\phi=\Tr \left[ F(\rho_{\phi,\eta},\vec{\Pi}_\phi)^{-1} \right]$ for the optimal phase measurement, while the dashed line represents the unattainable SLD-CRB $C^{\text{S}}_{\phi ,\eta}$.
\textbf{(a)} Relative difference
% $1-C^{\text{S}}_{\phi ,\eta}/C^{\text{H}}_{\phi ,\eta}$
between the SLD-CRB and the HCRB.
\textbf{(b)} Relative difference
% $1-C^{\text{H}}_{\phi ,\eta}/C^{\text{C}}_\phi$ 
between the HCRB and the classical CRB for the optimal phase measurement $C^{\text{C}}_\phi=\Tr \left[ F(\rho_{\phi,\eta},\vec{\Pi}_\phi)^{-1} \right]$; this quantity is zero (up to numerical noise) for $N\leq 6$.
}
\label{fig:HBstatesHCRB}
\end{figure}

More concretely, we study Holland-Burnett states, a family of states particularly resilient to imperfections~\cite{Datta2011}; we also fix $W = \id_2$.
Fig.~\eqref{fig:HBstatesHCRB} shows the classical CRB $C^\text{C}_\phi=\tr\left[F\left(\rho_{\phi,\eta},\vec{\Pi}_\phi\right)^{-1} \right]$, along with the HCRB (computed by solving the SDP numerically) and the SLD-CRB $C^{\text{S}}_{\phi ,\eta}$, as a function of $\eta$ for $N$ up to $14$.
Panel (a) shows that the HCRB is over 30\% tighter than the SLD bound, especially for intermediate transmissivities.
Panel (b) shows that the measurement we consider attains the HCRB for certain values of $N$ and $\eta$, i.e. the relative difference is zero up to numerical noise.
Even when the bound is not attained, the relative difference remains small at around 4\% for $N=14$.

For generic 1-photon states ($N=1$) we have found the analytical conditions for the HCRB to be attained by $\vec{\Pi}_\phi$.
For  $| \psi_\text{in} \rangle = c_0 | 0 ,  1 \rangle + c_1 | 1 ,  0 \rangle$ (with $|c_0|^2 +|c_1|^2 =1$) we have $C^\text{C}_\phi = C^\text{H} \left( \rho_{\phi,\eta} ; \id_2 \right)$ as long as $|c_1|^2 \geq 1/2$ or $\left(1- |c_1|^2/|c_0|^2 \right)/2 \leq \eta \leq 1 $.
The relative difference $1-C^\text{H}/C^\text{C}_\phi$ is at most 4.9\% and always zero for $\eta\geq 1/2$ (see Sec.~III.A of~\cite{SM} for details).
A numerical analysis on random states for higher values of $N$ suggests that there is indeed a threshold value of $\eta$, increasing with $N$, above which $\vec{\Pi}_\phi$ attains the HCRB.

Finally, we remark that working in the space $\mathcal{L}_\mathsf{h}^r(\mathcal{H})$ provides a distinct advantage for the numerics, since the Hilbert space dimension is $(N^2+3N+2)/2$, while $\rho_{\phi , \eta}$ has rank $r=N+1,$whereby $\tilde{d} =(N+1)^3 < (N+1)^4$.

% In these instances the HCRB is attainable at the single copy level and coincides with the most informative CRB (MICRB)~\cite{Nagaoka1989} $C^{\text{MI}} =\min_{\vec{\Pi}} \tr[ W F(\rho_{\vec{\theta}},\vec{\Pi})^{-1}]$, where the optimization is over all POVMs.
% % coincides with the MICRB, i.e. it  (again, this holds if the matrix $W$ is diagonal).
% For two-parameter models, Nagaoka introduced a further bound $C^{\text{N}}$~\cite{Nagaoka1989}, that sits between the HCRB and the MICRB: $C^{\text{MI}}_{\vec{\theta}} \geq C^{\text{N}}_{\vec{\theta}}  \geq C^{\text{H}}_{\vec{\theta}}$.
% For the problem at hand, we verified numerically that $C^{\text{N}}_{\vec{\theta}}  = C^{\text{H}}_{\vec{\theta}}$.

\emph{3D magnetometry}---Noiseless 3D magnetometry, another illustrative example of multi-parameter quantum metrology, has been studied terms of the QFIM~\cite{Baumgratz2015}.
Here, we highlight the necessity of using the HCRB for this problem and present results on 3D magnetometry using $M$ qubits in the presence of dephasing noise.
The parameters to be estimated $\vec{\varphi}=(\varphi_1,\varphi_2,\varphi_3)$ appear via the single-qubit Hamiltonian $H^{(j)}(\vec{\varphi}) = \vec{\varphi} \cdot \vec{\sigma}^{(j)}$, where $\vec{\sigma}^{(j)}$ is a vector of Pauli operators acting on the $j^{\text{th}}$ qubit. 
The parameters are imprinted on the probe state via the unitary $U_{\vec{\varphi}}=\bigotimes_j^{M} \exp\left(-\I H^{(j)}(\vec{\varphi})\right)$.
This is followed by local dephasing along the $z-$axis described by the single qubit CPTP map $ 2 \mathcal{E}_\gamma [ \rho ]= (1+\sqrt{1-\gamma}) \rho + (1-\sqrt{1-\gamma}) \sigma_z \rho \sigma_z$, with $\gamma \in [0,1]$; an approximation valid when the sensing time is short.

We use as probe states the family of 3D-GHZ states
\begin{equation}
\label{eq:3DGHZ}
\ket{\psi^{\text{3D-GHZ}}_M} = \frac{1}{\mathcal{N}}\sum_{k = 1}^3 \ket{\phi_k^+}^{\otimes M} + \ket{\phi_k^-}^{\otimes M} ,
\end{equation}
which was shown to present Heisenberg scaling in the noiseless case~\cite{Baumgratz2015}; $\ket{\psi_k^\pm}$ are the eigenvectors corresponding to the $\pm 1$ eigenvalues of the $k^{\text{th}}$ Pauli matrix and $\mathcal{N}$ is the normalisation.
The final state for which we compute the bound is $\rho_{\vec{\varphi}}= \mathcal{E}_{\gamma}^{\otimes M} [ U_{\vec{\varphi}} | \psi_M^{\text{3D-GHZ}} \rangle \langle \psi_M^{\text{3D-GHZ}} | U_{\vec{\varphi}}^\dag ]$; for the numerical results we choose equal parameter values $\varphi_i = 1 \, \forall i$ and $W=\id_3$.

\begin{figure}[t]
\includegraphics{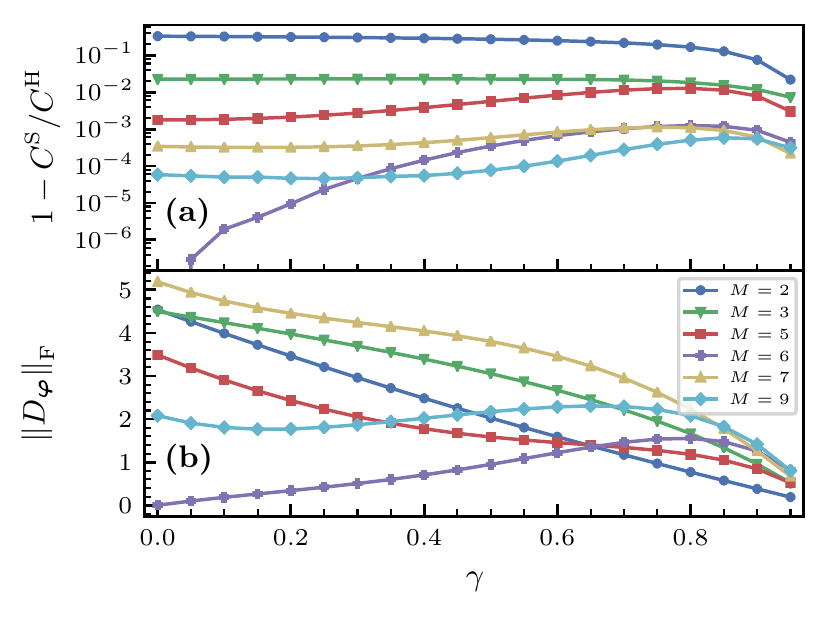}
\caption{
Comparison of the SLD-CRB, the HCRB (both with $W=\id_3$) and incompatibility for a $M$-qubit 3D-GHZ probe state undergoing a 3D phase rotation with equal values of parameters $\varphi_i=1$, followed by local dephasing of strength $\gamma$ in the $z$-direction.
\textbf{(a)} Relative difference $1-C^{\text{S}}/C^{\text{H}}$ between the SLD-CRB and HCRB.
\textbf{(b)} Incompatibility of the quantum statistical model, quantified by the Frobenius norm of the matrix $(D_{\vec{\varphi}})_{ij}=\Im \left( \Tr\left[ L_j L_i \rho_{\vec{\varphi}}\right] \right)$,
% i.e. $\Vert D_{\vec{\varphi}} \Vert_\text{F} = \sqrt{ 2 \left(| (D_{\vec{\varphi}})_{1,2} |^2 + | (D_{\vec{\varphi}})_{1,3} |^2 + | (D_{\vec{\varphi}})_{2,3} |^2 \right) }$,
showing incompatibility for all $\gamma$ (except for $M=6$ and $\gamma=0$).
Data for $M=4,8$ is not shown on the plot, since such models appear to be asymptotically classical, with $\Vert D_{\vec{\varphi}} \Vert_{\mathrm{F}} \lesssim 10^{-7}$ (consistently we find $1-C^{\text{S}}/C^{\text{H}} \lesssim 10^{-6} $).
}
\label{fig:MagComp}
\end{figure}

In Fig.~\eqref{fig:MagComp} we show the non-trivial relationship between the HCRB, the SLD-CRB and incompatibility for this quantum statistical model, as a function of the dephasing strength $\gamma$.
 % (and multiples of $8$).
We quantify the incompatibility of the model with the magnitude of the matrix $D_{\vec{\theta}}$, capturing the violation of the weak commutativity condition~\eqref{eq:weak_comm}; in particular, we use the Frobenius norm $\Vert D_{\vec{\varphi}}\Vert_{\text{F}}^2=\sum_{ij} | (D_{\vec{\varphi}})_{ij} |^2$.
Panel (a) shows that the relative difference $1-C^\text{S }/C^\text{H}$ is monotonically decreasing for 2 and 3 qubits, while it has a non-monotonic behaviour for 5 or more qubits.
Panel (b) shows that this behaviour is not always reflected at the level of incompatibility; this is remarkably different from the simple monotonic relationship found for 2-parameter pure state models~\cite{Matsumoto2002,Matsumoto1997}.
Furthermore, whilst the matrices $D_{\vec{\theta}}$ have a comparable magnitude for different number of qubits, the relative differences do not, e.g., being around 0.2\% for 5 qubits and around 30\% for 2 qubits.
% This is in contrast to what one would expect from the results of~\cite{Matsumoto2002}, where the difference between the 2-parameter HCRB and SLD-CRB for pure states is shown to be a monotonic function of the single independent element of $D_{\vec{\theta}}$.
% \textcolor{red}{\textsf{ Actually we should check what happens with the analytical result of~\cite{Suzuki2016a} for qubits for 2 parameters, to see if there is any somewhat similar "monotonic" behaviour.}}

Our SDP formulation grants us a previously inaccessible ease in the evaluation of the HCRB.
In turn, this enables us to get these glimpses into the non-commutative information geometry of 3-parameter mixed state problems of non-trivial dimension.
% Our SDP formulation of HCRB enables us to get these glimpses into the non-commutative information geometry of 3-parameter mixed state problems of non-trivial dimension.
From this example we see that the pair of matrices $J^\text{S}$ and $D_{\vec{\theta}}$ are not sufficient for a complete description of a given state's performance for multi-parameter estimation.

Finally, we concentrate on the noiseless ($\gamma=0$) case with two qubits, noting that a single qubit does not allow to estimate all the components of $\vec{\varphi}$~\cite{Baumgratz2015}.
In Fig.~\eqref{fig:MagComp} we see that for $M=2$ the SLD bound is considerably looser than the HCRB, with a relative difference around 30\% for $\gamma=0$.
On the contrary, we conjecture that the HCRB is attainable with single-copy projective measurements.
% we found that the HCRB is a very tight bound in this situation, and we conjecture that it 
We base this on the numerical equality between the HCRB and a numerical minimization of the classical scalar CRB over all 2-qubit projective measurements.
For 5000 random initial states with parameter values taken from five sets, the relative difference between the two quantities was always found to be smaller than $10^{-4}$ (see Sec.~IV.A of~\cite{SM} for details).
% ; this numerical boundary was chosen for the precision of the numerical global optimization over projective measurements.
% This result is obtained by comparing the HCRB with .
% To validate our finding % and $\comm{1000}$ random parameter values~\cite{SM}.
% (more details in the SM).
% This simple setup might prove amenable to experimental metrological applications.
While for pure states the HCRB is always attainable with single-copy measurements, the optimal POVM needs not be projective~\cite{Matsumoto2002}, making it harder to implement experimentally.
This finding shows that optimal protocols for 3D-magnetometry with two qubits may be not too far from experimental reach.
% Such a POVM is projective only if $\Im Z\left[ \vec{X}^{\text{opt}} \right] = 0$, where $\vec{X}^{\text{opt}}$ is the set of operators attaining the minimum in~\eqref{eq:HCRB}.
% For 2-qubit noiseless estimation we have that $\Im Z\left[ \vec{X}^{\text{opt}} \right] \neq 0$; nonetheless we always found a difference between the HCRB and the numerically optimized CRB of projective measurements smaller than \comm{$10^{-4}$}, which is comparable to the precision of the numerical optimization~\cite{SM}.

\emph{Conclusions}---We have shown how to evaluate the HCRB by solving a SDP, making it more easily accessible than previously believed.
This enabled us to study two examples---optical interferometry and 3D magnetometry---and gather numerical evidence that the HCRB is attainable by single-copy projective measurements, whereas the SLD bound is not.
% may unexpectedly be attained by a single-copy projective measurement.
%, both for phase and loss estimation and 3D magnetometry.
These findings suggest that there may be further unstudied cases where the HCRB is easier to attain than naively expected.
They also illustrate the potential of our formulation to enable new discoveries in multi-parameter quantum estimation, which should aid a deeper quantitative understanding of quantum measurements more generally.

% We have shown how to evaluate the HCRB by solving a SDP, making it a more accessible quantity to compute than previously believed.
% We foresee that this metrological bound will find application in many multi-parameter problems.
% We have shown some examples where the HCRB can unexpectedly be attained by a single-copy projective measurement, both for phase and loss estimation and 3D magnetometry.
% This finding suggests that there might be other unstudied cases where the HCRB is easier to attain than naively expected.
% This calls for a deeper quantitative understanding of quantum measurements.
% In particular, while rank-deficient density matrices are extremely relevant in physics, many mathematical results in quantum estimation are obtained for full-rank models or pure state models~\cite{Hayashi2005}.

% In the future it would be interesting to perform a detailed comparison between the HCRB and recently introduced metrological bounds based on ``digital'' information-theoretical quantities~\cite{Maccone2017,Peng2018}.
% Furthermore, since the HCRB takes into account more deeply the non-commutativity of observables, it would be interesting to use it to better explore the interplay between quantum optimal control and multi-parameter metrology~\cite{Yuan2016b,Liu2017d}.

\begin{acknowledgements}
\emph{Acknowledgements}---We thank E.~Bisketzi, D.~Branford, M.~G.~Genoni and A.~Saltini for many fruitful discussions.
% We thank Dominic Branford and Evangelia Bisketzi for many fruitful discussions.
This work has been supported by the UK EPSRC (EP/K04057X/2), the UK National Quantum Technologies Programme (EP/M01326X/1, EP/M013243/1) and Centre for Doctoral Training in Diamond Science and Technology (EP/L015315/1).
\end{acknowledgements}

\nocite{Liu2014a,Liu2014c,nielsen2010quantum,Shammah2018,Diamond2016,Agrawal2018,scs,ODonoghue2016}
\bibliography{HCRBbiblio}
\end{document}

% \appendix
\renewcommand\theequation{S\arabic{equation}}
\renewcommand\thefigure{S\arabic{figure}}
\renewcommand\thepage{SM-\arabic{page}}
\setcounter{equation}{0}
\setcounter{page}{1}
% \widetext
\title{Evaluating the Holevo Cramér-Rao bound for multi-parameter quantum metrology:\\Supplemental Material}
\author{Francesco Albarelli}
\email{francesco.albarelli@gmail.com}
\affiliation{Department of Physics, University of Warwick, Coventry CV4 7AL, United Kingdom}
\author{Jamie F. Friel}
\email{j.friel@warwick.ac.uk}
\affiliation{Department of Physics, University of Warwick, Coventry CV4 7AL, United Kingdom}
\affiliation{EPSRC Centre for Doctoral Training in Diamond Science and Technology, UK}
\author{Animesh Datta}
\email{animesh.datta@warwick.ac.uk}
\affiliation{Department of Physics, University of Warwick, Coventry CV4 7AL, United Kingdom}
\maketitle

This document is structured as follows.
In Sec.~\ref{app:ArbRank} we explain in details the structure of the equivalence class of observables induced by rank-deficient states and we introduce a basis for such a space built from the eigenvectors of the density matrix.
In Sec.~\ref{app:SDP} we explicitly show how to convert the convex optimization used to evaluate the Holevo Cramér-Rao bound (HCRB) to a SDP in a standard form.
In Sec.~\ref{app:phaseloss} we give more details on simultaneous phase and loss estimation in optical interferometry and we give details on the analytical results for the single-photon case.
Finally, in Sec.~\ref{app:3Dmag} we give more details about 3D magnetometry with multi-qubit systems and we present our numerical evidence for the attainability of the HCRB with single-copy projective measurements for 2-qubit noiseless systems.

\section{Space of observables for arbitrary-rank states}
\label{app:ArbRank}
We consider a density matrix $\rho_{\vec{\theta}}$ of rank $r$, with $1\leq r \leq d$.
We will now show that a non-trivial kernel of $\rho_{\vec{\theta}}$, i.e. $d<r$, gives rise to an equivalence class of hermitian matrices (observables) that produce the same results for scalar quantities, i.e., expectation values of polynomials of such observables.
The core of this argument is explained in Holevo's book~\cite[Sec. 2.10]{Holevo2011b}, 
where the equivalence class of square summable operators is introduced to deal with unbounded infinite-dimensional observables.
See also related discussions about SLDs in~\cite{Fujiwara1995} for pure states and in~\cite{Liu2014a,Liu2014c} for arbitrary rank states.
We consider a spectral decomposition of the density matrix:  
\begin{equation}
\rho = \sum_i^r p_i |\psi_i \rangle \langle \psi_i | = U_\rho \mathrm{diag} \left( p_1, \dots, p_r, 0,  \dots, 0 \right) U_\rho^\dag \,,
\end{equation}
in the sum we consider only strictly positive eigenvalues $p_i > 0$, so that $r$ is the rank of $\rho$.
The unitary matrix $U_\rho$ contains the eigenvectors $| \psi_i \rangle$ in the support of $\rho$ as the first $r$ columns and an orthonormal basis $| \phi_k \rangle$ for the kernel of $\rho$ as the remaining $d-r$ columns.
For later convenience we label the basis $| \phi_k \rangle$ starting from $r+1$ instead of $1$, i.e. $k=r+1, \dots, d$.

Let us study the structure of linear operators on $\mathcal{H}$, i.e. elements of the Hilbert space $\mathcal{L}\left(\mathcal{H}\right)$, in the basis of eigenvectors of $\rho$.
It is important to retain both the eigenstates in the support $|\psi_i \rangle$ and those in the kernel $|\phi_i \rangle$.
An arbitrary operator $A$ on a $d$-dimensional space has thus a block structure
\begin{equation}
A = 
\begin{pmatrix} A^\mathsf{s} & A^\mathsf{sk} \\
A^\mathsf{ks} & A^\mathsf{k} 
\end{pmatrix},
\end{equation}
where the labels $\mathsf{s}$ and $\mathsf{k}$ stand for support and kernel respectively.
In terms of matrix elements these blocks are defined as follows: $A^\mathsf{s}_{ij} = \langle \psi_i | A | \psi_j \rangle, \; A^\mathsf{k}_{kj} = \langle \phi_{k+r} | A | \phi_{j+r} \rangle, \; A^\mathsf{ks}_{il} = \langle \psi_i | A | \phi_{l+r} \rangle, \; A^\mathsf{sk}_{kj} = \langle \phi_{k+r} | A | \psi_j \rangle$, with $i,j \in [1,r]$ and $k,l \in [r+1,d]$.
Also the state $\rho$ after diagonalization has a block form, but only the upper diagonal block is non-zero: 
\begin{equation}
\rho = \begin{pmatrix} \mathrm{diag}(p_1, \dots, p_r) & \zero_{r,d-r} \\
\zero_{d-r,r}  & \zero_{d-r} \\
\end{pmatrix},
\end{equation}
where $\zero_{p,q}$ detones the rectangular $p \times q$ zero matrix and $\zero_{p}$ the square $p \times p$ square zero matrix.
For later convenience we also introduce the vector of non-zero eigenvalues $\vec{p} = (p_1, \dots, p_r)^\mathsf{T}$.
The expectation value of the operator is only determined by the projection on the support $A^\mathsf{s}$:
\begin{equation}
\mathrm{Tr} \left[ A \rho \right] = \mathrm{Tr}_\mathsf{s} \left[ A^\mathsf{s} \rho \right] \,.
\end{equation}
The expectation value of a product of operators (or powers of $A$) also involves off-diagonal terms
\begin{equation}
\label{eq:HermProduct}
\mathrm{Tr} \left[ A B \rho \right] = \mathrm{Tr}_\mathsf{s} \left[ \left( A^\mathsf{s} B^\mathsf{s} + A^\mathsf{sk} B^\mathsf{ks} \right) \rho \right] \neq \mathrm{Tr}_\mathsf{s} \left[ A^\mathsf{s} B^\mathsf{s}  \rho \right] \,,
\end{equation}
where now $A^\mathsf{sk} B^\mathsf{ks} \in \mathcal{L}\left(\mathrm{supp}(\rho)\right)$ is an operator acting only on the support of $\rho$.
If we restrict to Hermitian operators we need to keep track of a single off-diagonal block, since $A^\mathsf{ks}={A^\mathsf{sk}}^\dag$.
For example, in the case of a pure state ($r=1$) the off-diagonal rectangular matrix $A^\mathsf{ks}$ becomes a row vector. 
Derivatives of the density matrix have the projection on the kernel equal to zero
\begin{equation}
\partial_\lambda \rho = \begin{pmatrix} 
(\partial_\lambda \rho)^\mathsf{s} & (\partial_\lambda \rho)^\mathsf{sk} \\
(\partial_\lambda \rho)^\mathsf{ks} & \zero_{d-r} \\
\end{pmatrix}.
\end{equation}
The off-diagonal elements are non-zero is because the eigenvectors themselves can change by changing the parameter $\lambda$, while we are assuming that the rank of the state is not changing.

From the the previous discussion we see that the blocks $A^\mathsf{k},B^\mathsf{k}$ do not appear in the expectation value of the product of Hermitian operators $A B$; this can be generalized to polynomials of an arbitrary number of Hermitian operators: the blocks acting on the kernel of the density matrix are always irrelevant.
Therefore we can consider an equivalence class of Hermitian operators on $\mathcal{L}_\mathsf{h}(\mathcal{H})$ by disregarding such a block:
\begin{equation}
\label{eq:Ask}
A =
\begin{pmatrix} A^\mathsf{s} & A^\mathsf{sk} \\
A^\mathsf{ks} & \sim 
\end{pmatrix}.
\end{equation}
More formally, this means that we can work with elements of the quotient space
\begin{equation}
\label{eq:quot_space_rank}
\mathcal{L}^r_\mathsf{h}(\mathcal{H}) \equiv \mathcal{L}_\mathsf{h}(\mathcal{H}) / \mathcal{L}_\mathsf{h}\left(\mathrm{ker}(\rho_{\vec{\theta}})\right) \,.
\end{equation}
The dimension of this space (the so-called codimension) is equal to $\dim\left[ \mathcal{L}_\mathsf{h}(\mathcal{H}) \right] - \dim\left[ \mathcal{L}_\mathsf{h}(\mathrm{ker}(\rho_{\vec{\theta}})) \right] = d^2 - (d-r)^2 = 2 d \, r - r^2\equiv \tilde{d}$.
% we remark that this is very different from considering the space $\mathcal{L}_\mathsf{h}\left(\mathcal{H}/\mathrm{ker}(\rho_{\vec{\theta}})\right)$.

We stress that for computing scalar quantities, i.e. expectation values, we can work in the reduced space $\mathcal{L}^r_\mathsf{h}(\mathcal{H})$.
However, different operators belonging to the same equivalence class have different properties when considered as elements of the full space $\mathcal{L}_\mathsf{h}(\mathcal{H})$.
In particular, the commutativity of two operators cannot be established by considering only the quotient space $\mathcal{L}^r_\mathsf{h}(\mathcal{H})$, i.e. different operators belonging to the same equivalence class may or may not commute depending on the components in the subspace $\mathcal{L}_\mathsf{h}(\mathrm{ker}(\rho_{\vec{\theta}}))$. 
This argument is important with regards to the attainability of metrological bounds for pure states~\cite{Fujiwara1995,Matsumoto2002}.

A different way to look at this result is to notice that when the state has rank $r$ we can safely restrict to Hermitian operators of rank $r$.
This can be understood in terms of the spectral decomposition of the hermitian matrix $A$.
The eigenvalues of $A$ corresponding to the eigenvectors \emph{not} in the support of $\rho$ do not influence any scalar quantity, therefore they are arbitrary and we can always choose them to be zero so that $A$ has rank $r$.
Accordingly, the number of free real parameters in a rank-$r$ $d$-dimensional Hermitian matrix is $2 d \, r - r^2$; this corresponds to the free parameters in the $r \times r$ Hermitian block $A^\mathsf{s}$ plus the parameters in the the rectangular $r \times (d-r)$ block $A^\mathsf{sk}$, i.e. $r^2 + 2 r (d-r ) = 2 d \, r - r^2 \equiv \tilde{d}$.
% In particular a rank-$r$ quantum state is described by $2 d \cdot r - r^2- 1$ real parameters, due to normalization.

\subsection{Hermitian operator basis from eigenvectors}
For our numerical implementation we choose to work in the following basis $\left\{ \lambda_i \right\}_{i=1,\dots,\tilde{d}}$ for $\mathcal{L}^r_\mathsf{h}(\mathcal{H})$, orthonormal w.r.t. the Hilbert-Schmidt product, i.e. $\Tr\left[ \lambda_i \lambda_j \right] = \delta_{ij}$:
\begin{equation}
\label{eq:basislambda}
\begin{split}
	\lambda_i = &|\psi_i \rangle \langle \psi_i | \quad           i=1, \dots , r\\ 
	\lambda_{r+\left[(j-1)^2-(j-1)\right]/2+i} = &\frac{|\psi_i \rangle \langle \psi_j | + |\psi_j \rangle \langle \psi_i |}{\sqrt{2}} \quad i=1, \dots , j-1; \; \; j=2, \dots , r\\
	\lambda_{r+(r^2 - r)/2+\left[(j-1)^2-(j-1)\right]/2+i}=&\I \frac{|\psi_i \rangle \langle \psi_j | - |\psi_j \rangle \langle \psi_i |}{\sqrt{2} } \quad i=1, \dots , j-1; \; \; j=2, \dots , r\\
	\lambda_{r+(r^2 - r)+r(k-r-1)+i}=&\frac{|\psi_i \rangle \langle \phi_k | + |\phi_k \rangle \langle \psi_i |}{\sqrt{2}} \quad i=1, \dots , r; \;\; k=r+1, \dots , d \\
	\lambda_{r+(r^2 - r)+r(d-r)+r(k-r-1)+i}=&\I \frac{|\psi_i \rangle \langle \phi_k | - |\phi_k \rangle \langle \psi_i |}{\sqrt{2}} \quad i=1, \dots , r; \;\; k=r+1, \dots , d
\end{split}
\end{equation}
where we have $r$ basis elements defined by the first equation, $(r^2-r)/2$ defined by the third and  $(r^2-r)/2$ by the fourth, $r(d-r)$ defined by the fifth and $r(d-r)$ by the sixth, for a total of $\tilde{d}=r+(r^2-r)+2r(d-r)=2d r - r^2$ Hermitian matrices.
Furthermore in listing the basis elements in each of the five groups in~\eqref{eq:basislambda} we first iterate over $i$ and then over $j$ or $k$.
As in the previous section, $|\psi_i \rangle$ represents eigenvectors in the support and $| \phi_k \rangle$ eigenvectors in the kernel.
Crucially, this basis is \emph{parameter-dependent}, since it is built from the eigenvectors $| \psi_j \rangle$ of the density matrix; this has to be taken into account when taking derivatives.

\subsubsection{Vectorization}
For an arbitrary Hermitian operator $A$ we introduce the bijective vectorization operation $\mathrm{Vec}: \mathcal{L}_\mathsf{h}^r \left( \mathcal{H} \right) \mapsto \mathbb{R}^{\tilde{d}}$:
\begin{equation}
\mathrm{Vec}\left( A \right) \equiv \left( \Tr\left[ A \lambda_1 \right], \dots , \Tr\left[ A \lambda_{\tilde{d}}\right] \right)^\mathsf{T} .
\end{equation}
By choosing the basis given in~\eqref{eq:basislambda}, the components of $\mathrm{Vec}\left( A \right)$ can be written in terms of the blocks in~\eqref{eq:Ask}.
Using the ordering prescription for the basis elements given previously, the components of $\vec{a}=\mathrm{Vec}(A)$ are: the diagonal elements of $A_\mathsf{s}$, the real and imaginary part of the upper triangle above the diagonal of $A_\mathsf{s}$ and the real and imaginary part of $A_\mathsf{ks}$.
With the basis~\eqref{eq:basislambda} these matrices are vectorized in column major order.
More explicitly:
\begin{equation}
\begin{split}
	a_i =  &\langle \psi_i | A | \psi_i \rangle  \quad           i=1, \dots , r\\
	a_{r+\left[(j-1)^2-(j-1)\right]/2+i} = &\sqrt{2} \Re \left[ \langle \psi_i | A | \psi_j \rangle \right] \quad i=1, \dots , j-1; \; \; j=2, \dots , r\\
	a_{r+(r^2 - r)/2+\left[(j-1)^2-(j-1)\right]/2+i}=&\sqrt{2} \Im \left[ \langle \psi_i | A | \psi_j \rangle \right] \quad i=1, \dots , j-1; \; \; j=2, \dots , r\\
	a_{r+(r^2 - r)+r(k-r-1)+i}=&\sqrt{2} \Re \left[ \langle \psi_i | A | \phi_k \rangle \right] \quad i=1, \dots , r; \;\; k=r+1, \dots , d \\
	a_{r+(r^2 - r)+r(d-r)+r(k-r-1)+i}=&\sqrt{2} \Im \left[ \langle \psi_i | A | \phi_k \rangle \right] \quad i=1, \dots , r; \;\; k=r+1, \dots , d
\end{split}
\end{equation}

In this basis, the columns of the matrix $\vec{X}$ introduced in the main text correspond to $\mathrm{Vec} \left( X_i \right)$, while the columns of the matrix $\partial \vec{s}_{\vec{\theta}} / \partial \vec{\theta}$ correspond to $\mathrm{Vec} \left( \partial \rho_{\vec{\theta}}/{\partial \theta_i} \right)$.
This is a slight abuse of notation, since in this basis this is not the Jacobian matrix of the vector $\vec{s}_{\vec{\theta}}$; this would be the case if we worked with a parameter independent basis.	

\subsubsection{Matrix representation of the inner product}
We consider the inner product between two Hermitian operators is defined as\footnote{This is also known as the RLD inner product~\cite{Hayashi2017c}.}
(Eq.~(9) in the main text):
% (Eq.~\eqref{eq:RLDinner} in the main text):
\begin{equation}
\Tr \left[ A B \rho_{\vec{\theta}} \right] = \vec{a}^\mathsf{T} S_{\vec{\theta}} \vec{b},
\end{equation}
where the last equality holds after choosing a basis of Hermitian operators, i.e.
\begin{equation}
\vec{a} = \mathrm{Vec}(A) \quad [S_{\vec{\theta}}]_{ij} = \Tr \left[ \lambda_i \lambda_j \rho_{\vec{\theta}} \right]
\end{equation}

With the basis choice~\eqref{eq:basislambda} we can write the matrix representation of the inner product explicitly.
We can write it in three separate blocks:
\begin{equation}
S_{\vec{\theta}}= S_{\vec{\theta}}^{\mathsf{d}} \oplus S_{\vec{\theta}}^{\mathsf{s}} \oplus S_{\vec{\theta}}^\mathsf{sk} \,,
\end{equation}
where the first block is the diagonal part and simply corresponds to the eigenvalues:
\begin{equation}
S_{\vec{\theta}}^{\mathsf{d}}  = \mathrm{diag}\left( \vec{p} \right) \;.
\end{equation}
The second block can be rewritten in term of two vectors of sum and differences of eigenvalues as
\begin{equation}
S_{\vec{\theta}}^{\mathsf{s}}  = 
\begin{pmatrix} \mathrm{diag}\left( \vec{p}_{+} \right) & \I \, \mathrm{diag}\left( \vec{p}_{-} \right) 
 \\
-\I \, \mathrm{diag}(\vec{p}_{-})  & \mathrm{diag}\left( \vec{p}_{+} \right)
\end{pmatrix},
\end{equation}
where the components of the two $(r^2-r)/2$ dimensional vectors are
\begin{equation}
(p_{+})_{\left[(j-1)^2-(j-1)\right]/2+i} = \frac{p_i+p_j}{2} \qquad
(p_{-})_{\left[(j-1)^2-(j-1)\right]/2+i} =	\frac{p_i-p_j}{2}
\quad i=1, \dots , j-1; \; \; j=2, \dots , r\\
\end{equation}

\begin{equation}
S_{\vec{\theta}}^{\mathsf{sk}}  = 
\begin{pmatrix} \bigoplus_{i=1}^{d-r} \mathrm{diag}\left( \vec{p} \right) & -\I \bigoplus_{i=1}^{d-r} \mathrm{diag}\left( \vec{p} \right)
 \\
\I \bigoplus_{i=1}^{d-r} \mathrm{diag}\left( \vec{p} \right) & \bigoplus_{i=1}^{d-r} \mathrm{diag}\left( \vec{p} \right)
\end{pmatrix}
\end{equation}
From this representation it is easy to see that the right half of this last matrix is proportional to the left half, with coefficient $-\I$; therefore we have $\mathrm{rank}\left( S_{\vec{\theta}}^{\mathsf{sk}} \right) = r (d - r) $.
Since the other block of the whole matrix is full-rank, i.e. $\mathrm{rank} \left(S_{\vec{\theta}}^{\mathsf{d}} \oplus  S_{\vec{\theta}}^{\mathsf{s}} \right) = r^2$, globally we have $\mathrm{rank}\left( S_{\vec{\theta}} \right) = r d $.
%%%%
\section{Explicit conversion to SDP in standard inequality form}
\label{app:SDP}
Let us remark that the matrix inequality~(7) in the main text asserts the positive semi-definiteness of the \emph{complex Hermitian} matrix $V - Z[\vec{X}]$.
% Let us remark that the matrix inequality~\eqref{eq:HCRB} in the main text asserts the positive semi-definiteness of the \emph{complex Hermitian} matrix $V - Z[\vec{X}]$.
Therefore also the linear matrix inequality (LMI) in~(11) in the main text is a LMI in the complex Hermitian sense, but we can recast it as a LMI in the \emph{real symmetric} sense by doubling the dimension, i.e.
% Therefore also the linear matrix inequality (LMI) in~\eqref{eq:SDPform2} in the main text is a LMI in the complex Hermitian sense, but we can recast it as a LMI in the \emph{real symmetric} sense by doubling the dimension, i.e. 
\begin{equation}
\begin{pmatrix} V & \vec{X}^\mathsf{T} R_{\vec{\theta}}^\dag \\
R_{\vec{\theta}} \vec{X} & \id_{\tilde{r}} \end{pmatrix}\succeq 0 \iff 
\\ 
\begin{pmatrix} V & \vec{X}^\mathsf{T} \mathrm{Re} R_{\vec{\theta}}^\mathsf{T} & \zero_{n} & \vec{X}^\mathsf{T} \mathrm{Im} R_{\vec{\theta}}^\mathsf{T} \\
\mathrm{Re} R_{\vec{\theta}} \vec{X} & \id_{\tilde{r}} & -\mathrm{Im} R_{\vec{\theta}} \vec{X} & \zero_{\tilde{r}} \\
\zero_{n} & -\vec{X}^\mathsf{T} \mathrm{Im} R_{\vec{\theta}}^\mathsf{T} & V & \vec{X}^\mathsf{T} \mathrm{Re} R_{\vec{\theta}}^\mathsf{T}\\
\mathrm{Im} R_{\vec{\theta}} \vec{X} & \zero_{\tilde{r}} & \mathrm{Re} R_{\vec{\theta}} \vec{X} & \id_{\tilde{r}} 
\end{pmatrix} \succeq 0 \;,
\end{equation}
where the matrix on the r.h.s. of the second inequality is now real symmetric;	 
we also used the identities $\Im R_{\vec{\theta}}^\dag = - \Im R_{\vec{\theta}}^\mathsf{T}$ and $\Re R_{\vec{\theta}}^\dag = \Re R_{\vec{\theta}}^\mathsf{T}$.

Now it easy to show that the convex optimization~(11) can be recast as a SDP in \emph{inequality form}~\cite[p.168]{Boyd}
% Now it easy to show that the convex optimization~\eqref{eq:SDPform2} can be recast as a SDP in \emph{inequality form}~\cite[p.168]{Boyd}
\begin{equation}
\label{eq:standardSDPform}
\begin{aligned}
& \underset{v \in \mathbb{R}^{\tilde{k}} }{\text{minimize}} & & c^\mathsf{T} v\\
& \quad \text{subject to} & & \sum_{i=1}^{\tilde{k}} v_i F_i + G \preceq 0 \\ 
& & & A x = b \,
\end{aligned}\;,
\end{equation}
where now the $\tilde{k}$-dimensional vector collecting the variables to optimize contains both the $(n^2 + n)/2$ real independent components of $V$ and the $n \tilde{d}$ real components of $\vec{X}$:
\begin{equation}
v=\left(V_{1,1},V_{1,2},V_{2,2}, \dots , V_{n,n},(\vec{X})_{1,1},(\vec{X})_{2,1},(\vec{X})_{3,1} \dots, (\vec{X})_{\tilde{d},n}\right)^\mathsf{T} \in \mathbb{R}^{\tilde{k}}\,,
\end{equation}
with $\tilde{k}= (n^2 + n)/2 + n \tilde{d}$.
The constant matrices in the inequality are
\begin{equation}
F_i = -\frac{\partial} { \partial v_i} \begin{pmatrix} V & \vec{X}^\mathsf{T} \mathrm{Re} R_{\vec{\theta}}^\mathsf{T} & \zero_{n} & -\vec{X}^\mathsf{T} \mathrm{Im} R_{\vec{\theta}}^\mathsf{T} \\
\mathrm{Re} R_{\vec{\theta}} \vec{X} & \id_{\tilde{r}} & \mathrm{Im} R_{\vec{\theta}} \vec{X} & \zero_{\tilde{r}} \\
\zero_{n} & -\vec{X}^\mathsf{T} \mathrm{Im} R_{\vec{\theta}}^\mathsf{T} & V & \vec{X}^\mathsf{T} \mathrm{Re} R_{\vec{\theta}}^\mathsf{T}\\
\mathrm{Im} R_{\vec{\theta}} \vec{X} & \zero_{\tilde{r}} & \mathrm{Re} R_{\vec{\theta}} \vec{X} & \id_{\tilde{r}} 
\end{pmatrix}
\end{equation}
and
\begin{equation}
G = - \zero_n \oplus \id_{\tilde{r}} \oplus \zero_n \oplus \id_{\tilde{r}}.
\end{equation}
The matrix $A$ and the vector $b$ are determined by vectorizing the matrix equality in~(11) and only affect the $n \tilde{d}$ dimensional subvector of $v$ containing the components of $\vec{X}$.
% The matrix $A$ and the vector $b$ are determined by vectorizing the matrix equality in~\eqref{eq:SDPform2} and only affect the $n \tilde{d}$ dimensional subvector of $v$ containing the components of $\vec{X}$.
The vector $c \in \mathbb{R}^{\tilde{k}}$ depends on the weight matrix $W$ and its only non-zero components are the $(n^2 + n )/2$ pertaining to elements of $V$.
The SDP can be further manipulated and put in \emph{standard conic form} by first translating the equality constraints to pairs of inequalities.
Finally, we remark that the time complexity of solving an SDP is polynomial in the size of the matrices appearing in~\eqref{eq:standardSDPform}~\cite{Boyd}.
% We remark again that all these steps are not necessary to get a numerical solution, since there are powerful convex optimization modelling frameworks, such as CVX or YALMIP, that directly take care of this.
% In this SDP the number of real parameters is $\frac{N^2 +N}{2} + N (d^2 -1)$, while the square symmetric matrices have dimension $2(N+d^2-1)$.
% The time complexity of solving an SDP can be roughly evaluated as $ O(n^2 p^2)$~\cite[p. 618]{Boyd}, where $n$ is the number of variables and $p$ is the dimension of the matrices, considering that $N\leq d^2 -1$ in our case the dominating order is $O ( N^2 d^8 )$ and the worst possible scaling will be $O(d^{12})$.
% More precise estimates of time complexity can be given by considering a specific algorithm and taking into account the sparsity of the matrices.

\section{Phase and loss estimation with fixed photon number states}
\label{app:phaseloss}
We consider initial states with a fixed photon number $N$ across two modes of a Mach-Zendher interferometer:
\begin{equation}
\label{eq:fixedphnum}
| \psi_\text{in} \rangle = \sum_{k=0}^{N} c_k | k ,  N-k \rangle.
\end{equation}
The evolved state after a lossy phase shift on the arm corresponding to the first mode, characterized by a transmissivity $\eta$ and a phase $\phi$, has the following direct sum form
\begin{equation}
\label{eq:PhaseLossState}
\rho_{\phi,\eta} = \bigoplus_{k=0}^N p_l | \psi_l \rangle \langle \psi_l | ,
\end{equation}
where now the two parameters to estimate are $\vec{\theta}=\left( \phi , \eta \right)^\mathsf{T}$.
Each term in the sum is the state corresponding to $l$ lost photons:
\begin{equation}
|\psi_l \rangle = \frac{1}{\sqrt{p_l}} \sum_{k=l}^{N} c_k e^{\I k \phi } \sqrt{B_l^k} \, |k -l , N - k \rangle , 
\end{equation}
where $B_l^k = \binom{k}{l} \eta^{k-l} (1-\eta)^l$ and $p_l$ represents the probability of losing $l$ photons.
These states are also orthogonal $\langle \psi_l | \psi_m \rangle = \delta_{lm}$.
Due to the direct sum structure of the state, it is possible to obtain the SLDs $L_\phi$ and $L_\eta$ analytically, as well as the diagonal QFIM $J^\text{S}=\mathrm{diag}(J^\text{S}_{\phi \phi},J^\text{S}_{\eta \eta})$; see~\cite{Crowley2014} for more details.
The SLDs are easy to obtain, since they share the block structure of the state and the state is in each block is pure
\begin{equation}
\begin{split}
L_\phi &= \bigotimes_{l=0}^N 2 \left( |\partial_\phi \psi_l \rangle \langle \psi_l | + |\psi_l \rangle \langle \partial_\phi \psi_l |  \right) \\ 
L_\eta &= \bigotimes_{l=0}^N \left[ ( \partial_\eta \ln p_l ) |\psi_l \rangle \langle \psi_l | + 2 |\partial_\eta \psi_l \rangle \langle \psi_l | + 2 |\psi_l \rangle \langle \partial_\eta \psi_l | \right] .
\end{split}
\end{equation}
The QFIM and the FIM for the projective measurement on the eigenstates of the phase SLD can be obtained analytically~\cite{Crowley2014}:
\begin{equation}
\begin{split}
J^{\text{S}} &= \begin{pmatrix}
J^{\text{S}}_{\phi \phi} & 0 \\
0 &  J^{\text{S}}_{\eta \eta } 
\end{pmatrix} =
\begin{pmatrix}
4 \left( \sum_{k=0}^N k^2 |c_k|^2  - \sum_{l=0}^N \frac{ \sum_{k=l}^N k |c_k|^2 B^k_l} { \sum_{k=l}^N |c_k|^2 B^k_l } \right) & 0 \\ 
0 & \frac{ \sum_{k=0}^N k |c_k|^2 }{\eta (1-\eta)} 
\end{pmatrix} \\
F(\rho_{\phi, \eta}, \vec{\Pi}_{\phi} ) &= \begin{pmatrix}
J^{\text{S}}_{\phi \phi} & 0 \\
0 &   J^{\text{S}}_{\eta \eta } - \frac{1}{4 \eta^2}  J^{\text{S}}_{\phi \phi}
\end{pmatrix},
\end{split}
\end{equation}
from these expressions it is easy to see that measuring $\vec{\Pi}_\phi$ is suboptimal for estimating the parameter $\eta$.

% The Hilbert space of the problem has a direct sum structure and has dimension $(N^2+3N+2)/2$, while the state~\eqref{eq:PhaseLossState} has rank $N+1$.
% Therefore the restriction to the space $\mathcal{L}_\mathsf{h}^r(\mathcal{H})$ brings a relevant advantage for computing the HCRB.
The class of Holland-Burnett (HB) states considered in the main text is obtained by interfering two Fock states, i.e. the two-mode state $|N/2 , N/2 \rangle$ (for even $N$), on a balanced beam splitter.
This is the first part of a Mach-Zehnder interferometer.
The resulting state with $N$ total photons distributed in the two modes is
\begin{equation}
\label{eq:HBstates}
| \psi_{\mathsf{HB}} \rangle = \sum_{k=0 }^{N/2} \frac{\sqrt{(2k)! (N-2k)!}}{2^{\frac{N}{2}} k! \left(\frac{N}{2}-k \right)!} |2 k , N - 2 k \rangle.
\end{equation}

All the numerical results of the SDP for this application to phase and loss estimation were obtained in MATLAB using the YALMIP modelling framework for convex optimization~\cite{Lofberg2004}.

\subsection{Analytical results for 1 photon states}
\label{subsec:anal}
Here we consider a generic single-photon $N=1$ input state, so that the initial state~\eqref{eq:fixedphnum} is simply a qubit
\begin{equation}
\label{eq:qubit_initial}
|\psi_\text{in} \rangle = c_0 | 0 , 1 \rangle + c_1 | 1 , 0 \rangle,
\end{equation}
while the corresponding evolved state is a block-diagonal qutrit state with rank $2$
\begin{equation}
\rho_{\phi,\eta} = p_1 | 0 , 0 \rangle \langle 0 , 0 | \oplus  p_0 |\psi_0 \rangle \langle \psi_0 |
\end{equation}
with $p_1 = (1-\eta)|c_1|^2$, $p_0 = 1 - p_1 $ and $| \psi_0 \rangle = \frac{1}{\sqrt{p_0}} \left( c_0 | 0 , 1 \rangle + c_1 \sqrt{\eta} |1, 0 \rangle \right)$.
Furthermore, for $N=1$ it is also sufficient to consider real positive coefficients $c_0=\sqrt{1-c_1^2}$ and $c_1 = |c_1| \in [0,1]$, since any relative phase between the two components does not play any role.
% This choice does not make the results less general, since the informational content of any state is invariant under parameter-independent unitaries $U$.
% The optimal operators $\vec{X}$ to evaluate the HCRB of the model described by the state $U \rho U^\dag$ and derivatives $\partial_i (U \rho U^\dag) = U \partial_i \rho U^\dag$ are obtained by applying the same unitary to the optimal operators for the state $\rho$ and the bound is thus invariant.

% We can obtain the solution of the HCRB by an explicit minimization of the function~\eqref{eq:HolFunExpl}, which becomes particularly simple for a two-parameter problem.
We can obtain the solution of the HCRB by an explicit minimization of the function~(13) in the main text, which becomes particularly simple for a two-parameter problem.
The term involving the the imaginary part of $Z \left[ \vec{X} \right]$ reduces to the absolute value of the only independent element of the skew-symmetric $2{\times}2$ matrix $\Im Z\left[ \vec{X} \right]$:
\begin{equation}
\begin{split}
h_{\phi,\eta}\left(X_\phi,X_\eta\right)=& \Tr \left[ X_\phi^2 \rho_{\phi,\eta} \right] + \Tr \left[ X_\eta^2 \rho_{\phi,\eta} \right] + 2 \left\lvert \Im \left( \Tr \left[ X_\phi X_\eta \rho_{\phi,\eta} \right] \right) \right\rvert\\
=& \begin{cases}
\Tr \left[ X_\phi^2 \rho_{\phi,\eta} \right] + \Tr \left[ X_\eta^2 \rho_{\phi,\eta} \right] \quad &\text{for } \quad \Im \left( \Tr \left[ X_\phi X_\eta \rho_{\phi,\eta} \right] \right) = 0  \\
\Tr \left[ X_\phi^2 \rho_{\phi,\eta} \right] + \Tr \left[ X_\eta^2 \rho_{\phi,\eta} \right] + 2 \Im \left( \Tr \left[ X_\phi X_\eta \rho_{\phi,\eta} \right] \right) \quad &\text{for } \quad \Im \left( \Tr \left[ X_\phi X_\eta \rho_{\phi,\eta} \right] \right) > 0  \\
\Tr \left[ X_\phi^2 \rho_{\phi,\eta} \right] + \Tr \left[ X_\eta^2 \rho_{\phi,\eta} \right] - 2 \Im \left( \Tr \left[ X_\phi X_\eta \rho_{\phi,\eta} \right] \right) \quad &\text{for } \quad \Im \left( \Tr \left[ X_\phi X_\eta \rho_{\phi,\eta} \right] \right) < 0 ,
\end{cases}
\end{split}
\label{eq:holfunonephot}
\end{equation}
where for simplicity we work only with $W=\id_2$.
Each operator $X_{\phi / \eta}$ is described by $2{\times }3{\times}2 - 4 = 8 $ real parameters, by working in the space of observables $\mathcal{L}_\mathsf{h}^2 ( \mathbb{C}^{3\times3} )$ for three dimensional density matrices with rank 2.
The orthonormal eigenbasis is given by the eigenvectors $\left\{ |0,0\rangle, |\psi_0\rangle , \frac{|\partial_\eta \psi_0 \rangle}{\sqrt{\langle \partial_\eta \psi_0 | \partial_\eta \psi_0 \rangle}} \right\}$, corresponding to the eigenvalues $\left\{ 1-p_0, p_0, 0 \right\}$.
The constraints $\Tr\left[ X_\phi \rho_{\phi,\eta} \right]=\Tr\left[ X_\eta \rho_{\phi,\eta} \right]=0$, $\Tr\left[ X_\phi \partial_\phi \rho_{\phi,\eta} \right]=1$, $\Tr\left[ X_\phi \partial_\eta \rho_{\phi,\eta} \right]=0$, $\Tr\left[ X_\eta \partial_\eta \rho_{\phi,\eta} \right]=1$ and $\Tr\left[ X_\eta \partial_\phi \rho_{\phi,\eta} \right]=0$ give rise to simple linear equalities that can be inverted explicitly to reduce the total number of free parameters from 16 to 10.

For the second and third lines of the piecewise function~\eqref{eq:holfunonephot} the gradient w.r.t. to the elements of $X_{\phi/\eta}$ is a linear function in those variables.
Since we have proven that this function is convex we already know that every local minimum is also a global one and we only need to find the values that make the gradient vanish.
It is not hard to check that the second line does not give any contribution, since the gradient of $\Tr \left[ X_\phi^2 \rho_{\phi,\eta} \right] + \Tr \left[ X_\eta^2 \rho_{\phi,\eta} \right] + 2 \Tr \left[ X_\phi X_\eta \rho_{\phi,\eta} \right]$ is never zero in the region $\Tr \left[ X_\phi X_\eta \rho_{\phi,\eta} \right] > 0$.
On the contrary, the global minimum is given by the piece of the function defined on the third line when $0 < c_1 < \frac{1}{\sqrt{2}} \, \land \, 0 < \eta < \frac{1}{2}\left(1- \frac{c_1^2}{1-c_1^2} \right)$.

The remaining case $\Im \left( \Tr \left[ X_\phi X_\eta \rho_{\phi,\eta} \right] \right) = 0$ is more complicated and it has to be treated as a constrained optimization problem.
Instead of solving the problem explicitly, it is easier to show that the measurement $\vec{\Pi}_\phi$ is optimal in this regime.
The measurement $\vec{\Pi}_\phi=\left\{ |l_1 \rangle \langle l_1 | , |l_2 \rangle \langle l_2 | , |l_3 \rangle \langle l_3 |\right\}$ is a projective measurement on the eigenstates $|l_i \rangle$ of the phase SLD $L_\phi = \sum_{i=1}^3 l_i |l_i \rangle \langle l_i | $.
The corresponding commuting Hermitian operators are given by 
\begin{equation}
X^*_\phi = \frac{1}{F(\rho_{\phi,\eta},\vec{\Pi}_\phi)_{\phi,\phi}} \sum_{i=1}^3 \frac{\langle l_i | \partial_\phi \rho_{\phi,\eta} | l_i \rangle}{\langle l_i |  \rho_{\phi,\eta} | l_i \rangle} | l_i \rangle \langle l_i | \qquad X^*_\eta = \frac{1}{F(\rho_{\phi,\eta},\vec{\Pi}_\phi)_{\eta,\eta}} \sum_{i=1}^3 \frac{\langle l_i | \partial_\eta \rho_{\phi,\eta} | l_i \rangle}{\langle l_i |  \rho_{\phi,\eta} | l_i \rangle} | l_i \rangle \langle l_i | ;
\end{equation}
these are observables with zero mean and covariance matrix equal to the inverse of the classical FIM, i.e. $\Tr\left[ X^*_i  X^*_j \rho_{\phi,\eta}\right]=\left[ F(\rho_{\phi,\eta},\vec{\Pi}_\phi)^{-1}\right]_{i,j}$.
One can check explicitly that the gradient of the function $\Tr \left[ X_\phi^2 \rho_{\phi,\eta} \right] + \Tr \left[ X_\eta^2 \rho_{\phi,\eta} \right]$  constrained to satisfy $\Im \left( \Tr \left[ X_\phi X_\eta \rho_{\phi,\eta} \right] \right)= 0$ is zero for $X_{\phi/\eta}=X^*_{\phi/\eta}$ and in particular it is a minimum.
For $c_1 \geq \frac{1}{\sqrt{2}} \, \lor \, \frac{1}{2}\left(\frac{1-2 c_1^2}{1-c_1^2} \right) \leq \eta \leq 1 $ the global minimum of $h_{\phi,\eta}\left(X_\phi,X_\eta\right)$ has to lie on this constrained region, since there are no minima in the other two pieces of the function; thus, in this regime the projective measurement $\vec{\Pi}_\phi$ attains the HCRB.

The final result is
\begin{equation}
C^\text{H}_{\phi,\eta}=
\begin{cases}
\frac{\left[ c_1^2 (\eta -1)+1\right] \left[ 4 \left(1-c_1^2\right) (1-\eta) \eta +1\right]}{4 c_1^2 \left(1-c_1^2\right) \eta } = \Tr \left[ F(\rho_{\phi,\eta},\vec{\Pi}_\phi)^{-1} \right] \; &\text{for } \, c_1 \geq \frac{1}{\sqrt{2}} \, \lor \, \frac{1}{2}\left(1- \frac{c_1^2}{1-c_1^2} \right) \leq \eta \leq 1 \\ 
\frac{1 +3 \eta  -4 \eta^3}{4 c_1^2 \eta } \; &\text{for }\,  0 < c_1 < \frac{1}{\sqrt{2}} \, \land \, 0 < \eta < \frac{1}{2}\left(1- \frac{c_1^2}{1-c_1^2} \right).
\end{cases}
\end{equation}

\begin{figure}[ht!]
\centering
\subfloat[]{\label{fig:reldiffRandomA}
\includegraphics[width=0.45\textwidth]{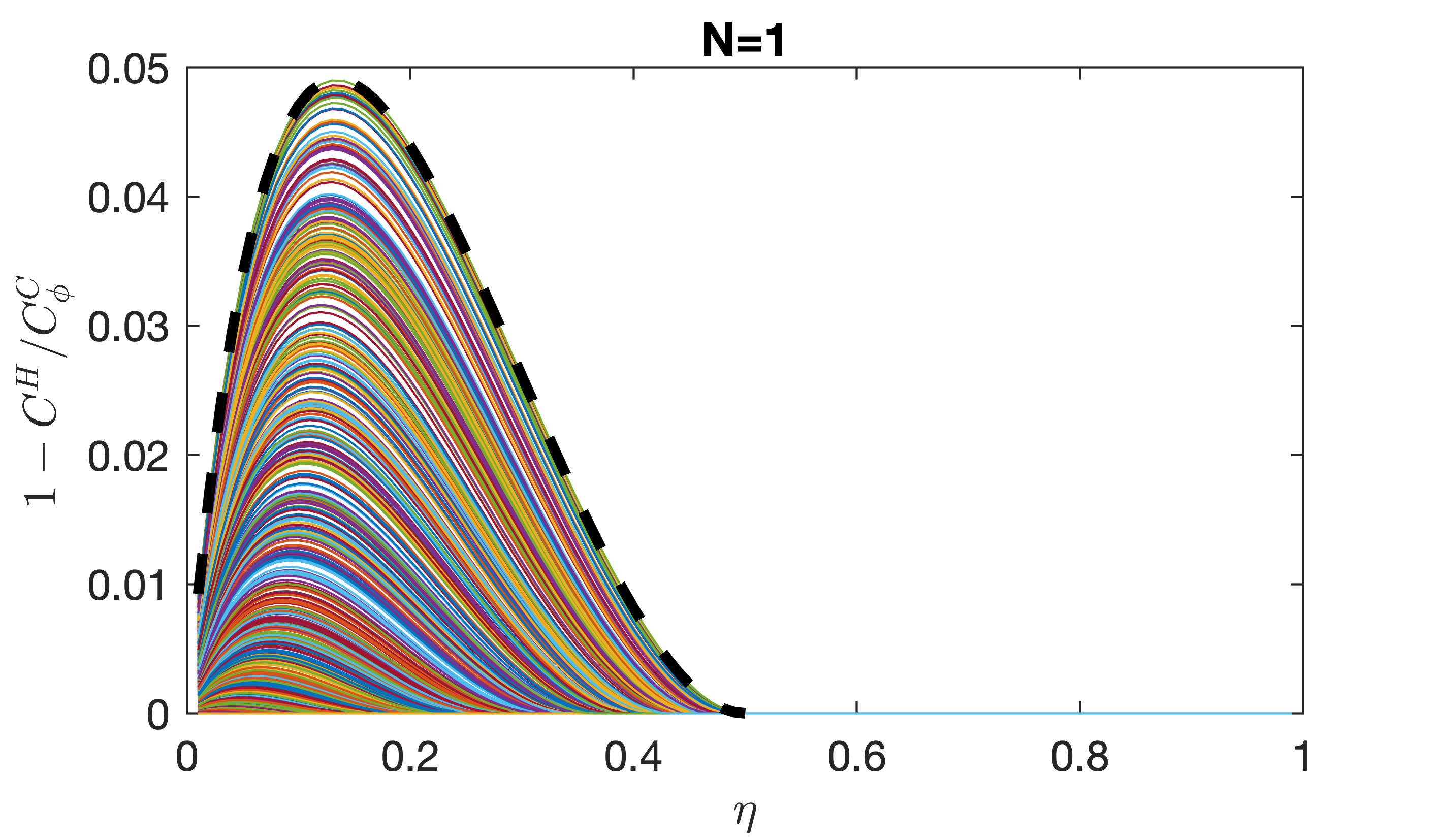}
}
\subfloat[]{\label{fig:reldiffRandomB}
\includegraphics[width=.45\textwidth]{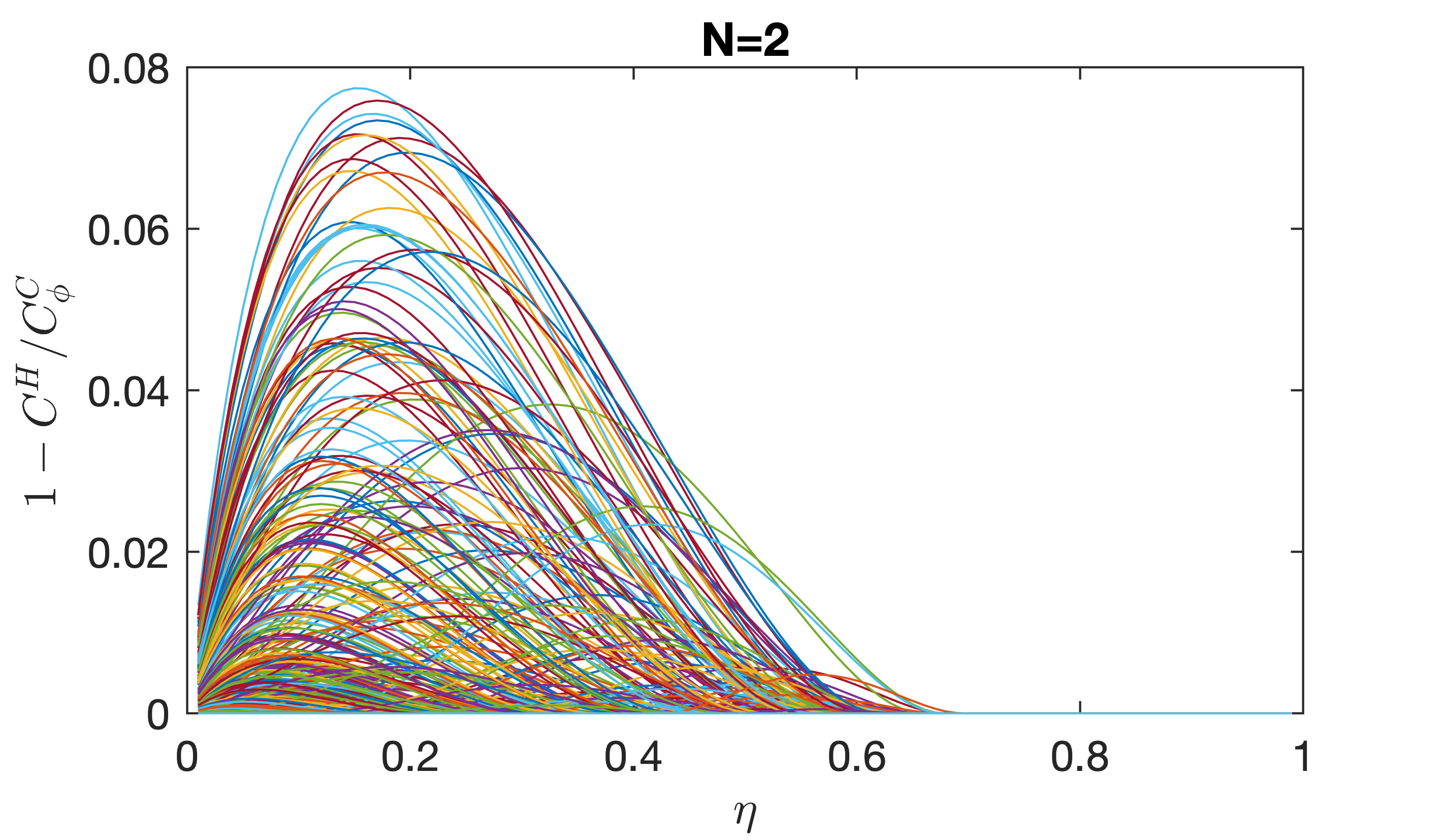}
}\\
\subfloat[]{\label{fig:reldiffRandomC}
\includegraphics[width=.45\textwidth]{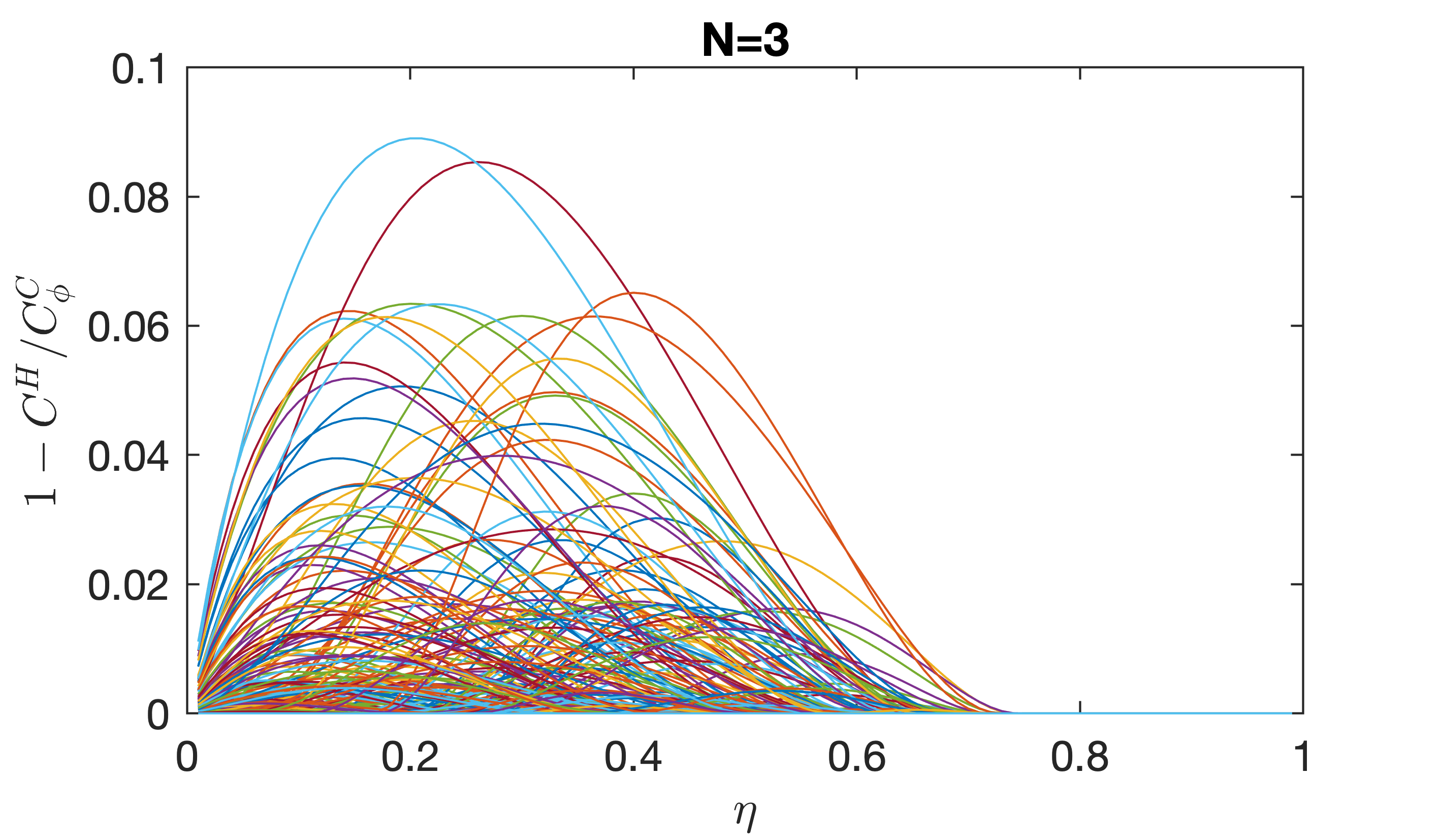}
}
\caption{Relative difference between the HCRB $C^{\text{H}}_{\phi ,\eta}$ (for $W=\id_2$) and the classical CRB $\Tr \left[ F(\rho_{\phi,\eta},\vec{\Pi}_\phi)^{-1} \right]$ pertaining to an optimal phase measurement.
The plots show this quantity computed for random initial states, as a function of the transmissivity $\eta$.
Panel (a) corresponds to $1000$ randomly generated initial pure states with $N=1$ total photon; the dashed thick black line shows the analytical result for the relative difference optimized over the coefficients of the initial state.
Panel (b) and (c) correspond to $1000$ randomly generated initial pure states with $N=2$ and $N=3$ respectively.
}\label{fig:reldiffRandom}
\end{figure}

We can study the relative difference between the HCRB and the classical CRB, i.e. the quantity $1-C^\text{H}_{\phi,\eta}/{\Tr \left[ F(\Pi_\phi)^{-1} \right]} $.
In Fig.~\ref{fig:reldiffRandom} we show this quantity for randomly generated initial states for $N=1,2,3$ as a function of $\eta$.
For $N=1$ we can maximize this quantity over the parameter $c_1$ characterizing the initial state, to obtain the maximal relative difference as a function of the efficiency:
\begin{equation}
\label{eq:maxline}
\max_{ c_1 } \left( 1-\frac{C^\text{H}_{\phi,\eta}}{\Tr \left[ F(\Pi_\phi)^{-1} \right]} \right) = 
\begin{cases}
0 \; &\text{for } \frac{1}{2} \leq \eta \leq 1 \\
\frac{\eta(1 - 2 \eta)^2}{1 + 4 \eta - 4 \eta^2} \; &\text{for } 0 < \eta < \frac{1}{2}.
\end{cases}
\end{equation}
The maximum relative difference is obtained in the limit $c_1 \to 0$: while both CRBs diverge in the limit of vanishing single photon components their ratio tends to a finite value.
The quantity in~\eqref{eq:maxline} is shown as a thick black dashed line in Fig.~\eqref{fig:reldiffRandomA}.
%%%%%%%
\section{3D magnetometry}
\label{app:3Dmag}
As described in the main text, we consider an initial pure input state on which we act first with a unitary parametrised by $\vec{\varphi}=\left( \varphi_x, \varphi_y, \varphi_z \right) \in \mathbb{R}^3$ and then with a dephasing channel along $z$, with a dephasing strength $\gamma \in [0,1]$.
The final state is represented by the following density matrix
\begin{equation}
\label{eq:finalrho3Dqubits}
\rho_{\vec{\varphi}} = \sum_{i=0}^{2^M-1} E_{\gamma,i} U_{\vec{\varphi}} | \psi_0 \rangle \langle \psi_0 | U_{\vec{\varphi}}^{\dag} E_{\gamma,i}^{\dag},
\end{equation}
here written in terms of the Kraus representation of the $M$-qubit dephasing channel. 
An alternative set of Kraus operators for the single qubit dephasing channel $\mathcal{E}_\gamma [ \rho ]$ introduced in the main text is~\cite{nielsen2010quantum}
\begin{align}
E_0 = \begin{pmatrix}
1 & 0 \\
0 & \sqrt{1-\gamma}
\end{pmatrix}, \qquad
E_1 = \begin{pmatrix}
0 & 0 \\
0 & \sqrt{\gamma}
\end{pmatrix}.
\end{align}
The $M$-qubit Kraus operators appearing in~\eqref{eq:finalrho3Dqubits} are labelled by $j \in [0, 2^N -1]$ and the binary representation of $j$ contains the information on which local dephasing operator acts on each qubit, e.g. $E_{\gamma,3}=E_{0} \otimes \dots  \otimes E_{0} \otimes E_{1} \otimes E_{1}$.
The $M$-qubit dephasing channel has the effect of contracting the Bloch sphere of each qubit to the $z$ axis, or equivalently to reduce the magnitude of the off-diagonal elements $| i\rangle \langle j |$ by a factor of $(1-\gamma)^{h(i,j)}$, where $h(i,j)$ is the Hamming distance between the binary numbers $i$ and $j$.
% by the Cartesian product of the set $\left\lbrace E_0,E_1 \right\rbrace$ with itself $N$ times.
The unitary dynamics is generated by the Hamiltonian $H = \sum_{k=1}^3 \varphi_k S_k  = \sum_{k=1}^3 \varphi_k \sum_{j=1}^M \sigma_k^{(j)}$, where we have introduced the global Pauli operators $S_k = \sum_{j=1}^M \sigma_k^{(j)} $.
The partial derivatives of the state can be expressed as
\begin{align}
\frac{\partial \rho_{\vec{\varphi}}}{\partial \varphi_k} = \I \sum_{i=0}^{2^M - 1} E_{i,\gamma} U_{\vec{\varphi}}\left[ | \psi_0\rangle \langle \psi_0 |,A_k \right] U_{\vec{\varphi}}^\dag E_{i,\gamma}^\dag,
\end{align}
in terms of the non-commuting generators $A_k$~\cite{Baumgratz2015}
\begin{equation}
A_k = \int_0^1 e^{\I \alpha H} S_k e^{-\I \alpha H} \text{d} \alpha.
\end{equation}
% In Fig.~\ref{fig:manypara} we show that the same qualitative behaviour presented in the main text for parameter values close to zero, i.e. $\varphi_x = \varphi_y = \varphi_z = 10^{-5} $ holds also in other regimes of parameter values.
% In particular, we present results for two qubits and we consider three more cases:
%  $\varphi_x = \varphi_y = \varphi_z = 1 $, $\varphi_x = 0$ and $\varphi_y = \varphi_z = 1 $,  $\varphi_x = \varphi_y = 0$ and $\varphi_z = 1 $.
% Whilst there are some differences in the rate of decrease, the relative difference is always monotonically decreasing towards 0.
% In fig~\ref{fig:absHol} we show the trends of the HCRB directly compared to the SLD-CRB.

Crucially, the dynamics we are considering is invariant for any permutation of the local qubit subsystems.
If we also choose a permutationally invariant initial state, such as the 3D-GHZ state, this symmetry can be used to reduce the size of the Hilbert space from exponential to polynomial in $M$.
To achieve this in our code for numerical calculations, we have taken advantage of the Python library for permutationally invariant two-level quantum systems introduced in~\cite{Shammah2018}.
Moreover, for the numerical implementation of our SDP we have employed the Python convex modelling framework CVXPY~\cite{Diamond2016,Agrawal2018} together with the solver SCS~\cite{scs,ODonoghue2016}.
With this SDP formulation we were able to obtain the data points shown in the main text on a desktop computer in a reasonable amount of time (each data point for $N=9$ using the permutationally invariant basis was obtained in less than half an hour).

% It is also worth while to note here that, while the 4-qubit states are asymptotically classical, it performs worse than 3 and 5 qubits in absolute terms over noise, this is one of the subtitles of the geometry of states at small number of qubits.
% \begin{figure}[h]
% \includegraphics[width = \textwidth]{figures/3para_abs.png}
% \caption{HCRB and SLD CRB over noise over 2-5 qubits}
% \label{fig:absHol}
% \end{figure}
% \begin{figure}[h]
% \includegraphics[width = \textwidth]{figures/2_qubit_diffpara_noise.png}
% \caption{
% Relative difference between HCRB and SLD-CRB as a function of noise for an initial two-qubit 3D-GHZ state $| \psi_2^{\text{3D-GHZ}} \rangle$ and for different values of the parameters to estimate.
% \comm{Show data points + add differences for different curves}
% }
% \label{fig:manypara}
% \end{figure}

\subsection{Attainability with projective measurements for two-qubit noiseless estimation}
Here we present our numerical evidence that the HCRB for noiseless estimation with $\vec{\varphi}$ with 2-qubit systems is attainable with projective measurements.
We parametrize a generic unitary of the 2-qubit system as
\begin{equation}
V_{\vec{x}} = \exp\left[ -\I \sum_{i,j=0}^{3} x_{ij} \lambda_i \otimes \lambda_j \right] 
\end{equation}
where we have introduced the Pauli basis of Hermitian operators for the two-qubit subsystems $\vec{\lambda} = \left( \id_2,\sigma_x, \sigma_y, \sigma_z \right)$ and the set of real coefficients $\vec{x} = \left\lbrace x_{ij} \right\rbrace_{i,j=0,\dots,3}$.
The eigenvectors of $V_{\vec{x}}$ are orthonormal and introduce a projective measurement $\vec{\Pi}_{\vec{x}}$ on the two-qubit system.
The classical Fisher information matrix $F\left( \ket{\psi}_{\vec{\varphi}},\vec{\Pi}_{\vec{x}} \right)$ associated to such a measurement on the evolved two-qubit state $\ket{\psi}_{\vec{\varphi}}$ is defined in Eq.~(4) in the main text.
% The classical Fisher information matrix $F\left( \ket{\psi}_{\vec{\varphi}},\vec{\Pi}_{\vec{x}} \right)$ associated to such a measurement on the evolved two-qubit state $\ket{\psi}_{\vec{\varphi}}$ is defined in Eq.~\eqref{eq:CFIM} in the main text.

We define the optimal projective scalar CRB (for $W=\id_3$) as 
\begin{equation}
\label{eq:optprojCRB}
C^{\text{proj}}_{\vec{\varphi}}= \min_{\vec{x}} \Tr \left[ F\left( \ket{\psi}_{\vec{\varphi}},\vec{\Pi}_{\vec{x}} \right)^{-1} \right]
\end{equation}
and we evaluate this quantity numerically.
The set of projectors is not convex and thus this has to be treated as a global optimization problem, which is general much harder to solve numerically than convex problems.
Nevertheless, the Nead-Medler algorithm for local optimization implemented in the Python's \verb|scipy| library gives good results by trying a few different random starting points for the parameter $\vec{x}$.
% We employ the \verb|scipy| implementation of Nead-Medler optimisation algorithm.
% The set of projectors is not convex and we found that standard gradient methods would almost always fail to converge to a global minimum.
% As it stands, we have no useful heuristics to aid the numerical optimisation.
% and this could be another interesting research question in order to make CFI optimisation more accessible numerically. 

% As mentioned in the main text, this is not in contradiction to the results of Matsumoto, but it does highlight the important distinction between attainability of matrix bounds and attainability of scalar bounds.
% As in Appendix~\ref{app:phaseloss} this pin points another gap in the theory of attainability of scalar bound, not only of rank deficient states but also pure states. 

\begin{figure}[h]
\includegraphics{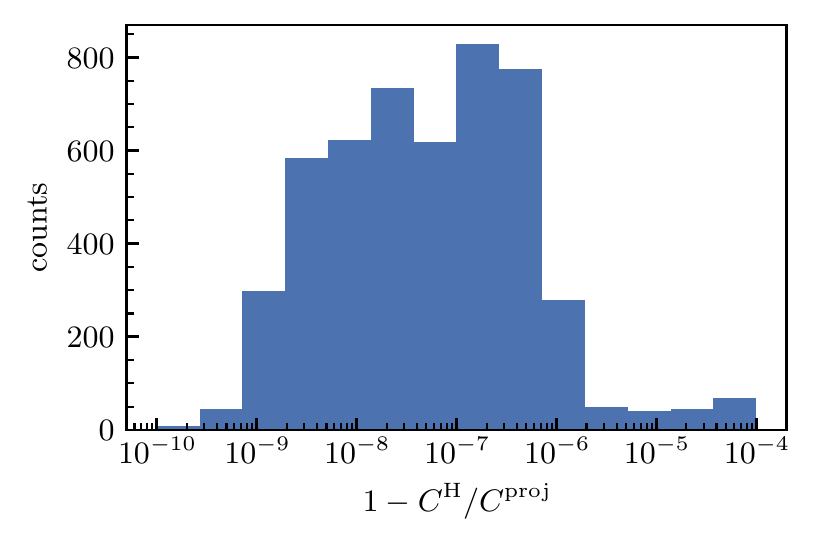}
\caption{
Histogram of the relative difference between the HCRB and the classical scalar CRB optimised over projective measurements for 2-qubit noiseless 3D magnetometry.
Data is obtained by generating random 2-qubit states and evolving them with parameter values $\vec{\varphi}_k$ for $k\in [1,5]$.
}
\label{fig:2qubitproj}
\end{figure}

In Fig.~\eqref{fig:2qubitproj} we show a histogram of the values obtained for the relative difference $1-C_{
\vec{\varphi}}^{\text{H}}/C^{\text{proj}}_{\vec{\varphi}}$ by generating 5000 random initial two-qubit pure states according to the Haar measure.
We tested five sets of parameter values $\vec{\varphi}_1 = (0,0,10^{-4})$, $\vec{\varphi}_2 = (0,1,1)$, $\vec{\varphi}_3 = (0,0,1)$, $\vec{\varphi}_4 =(1 , 1 , 1)$, $\vec{\varphi}_5 =(0.3305 , 1.6584 , 0.4844)$, each with 1000 random initial states; all data is plotted in the same histogram.~\footnote{Further checks with other parameter values did not produce different results.}

The main empirical conclusion is that we were able to obtain relative differences smaller than $10^{-4}$ for all states, while the vast majority of states we got a relative difference smaller than $10^{-5}$.
We remark that $10^{-4}$ is a threshold that we set for the global optimization but we believe that it could be lowered at the expense of an increased computational cost.

This is only numerical evidence and in principle there could be a very small but finite difference between the two bounds so that they do not actually coincide.
We have not been able to formally prove the equivalence, but the question of attainability of the HCRB with single-copy projective measurements is an important one that we plan to explore more in detail in future works.
Nonetheless, these numerical results show that the difference between the HCRB and the optimal projective bound is for all practical purposes negligible.

\bibliography{HCRBbiblio}